\documentclass[pre,twocolumn,amsmath,amssymb,nofootinbib,floatfix,superscriptaddress]{revtex4}

\usepackage{graphicx,bm,xcolor}
\usepackage{enumerate}
\usepackage{graphicx,bm}
\usepackage[normalem]{ulem}

\makeatletter
\def\graphicscale{\twocolumn@sw{0.3}{0.4}}
\def\graphicthreescale{\twocolumn@sw{0.3}{0.4}}

\begin{document}

\title{Lattice spin systems coupled to a compact U(1) gauge field\\
with negative gauge coupling favoring frustration}

\author{Claudio Bonati} 
\affiliation{Dipartimento di Fisica dell'Universit\`a di Pisa
        and INFN Sezione di Pisa, Largo Pontecorvo 3, I-56127 Pisa, Italy}

\author{Ettore Vicari} 
\affiliation{Dipartimento di Fisica dell'Universit\`a di Pisa,
        Largo Pontecorvo 3, I-56127 Pisa, Italy}

\date{\today}

\begin{abstract}
We investigate the phase diagram and nature of the phase transitions
of three-dimensional compact lattice Abelian Higgs models with
negative gauge coupling. These models describe multicomponent complex
scalar fields coupled to a compact U(1) gauge field. Negative gauge
coupling favors frustrated plaquette configurations, competing with
the tendency of the matter ferromagnetic-like interaction to suppress
frustration. Monte Carlo simulations and finite-size scaling analyses
of models with one- and two-component scalar fields of unit charge
reveal a substantially richer phase structure than at nonnegative
gauge coupling. For two components, the O(3) transition line
associated with gauge-invariant bilinear ordering extends into the
negative-coupling region, and turns into a first-order transition line
with decreasing $K$, toward the fully-frustrated $K\to -\infty$
limit. An analogous first-order transition line is also found for the
one-component scalar matter with negative gauge coupling.  Both one-
and two-component models exhibit two additional continuous transition
lines at sufficiently large matter coupling. Their energy-cumulant
scaling is consistent with three-dimensional Ising and XY critical
exponents, suggesting topological transitions in the $\mathbb Z_2$ and
XY gauge universality classes, respectively. These results uncover
additional phases generated by the competition between matter ordering
and dynamical gauge frustration, whose physical characterization
remains an open question.
\end{abstract}

\maketitle


\section{Introduction}
\label{intro}

Despite their apparent simplicity, spin models with short-ranged
interactions can display a remarkably rich variety of phases and phase
transitions, arising from the competition between local interactions,
thermal or quantum fluctuations, and symmetry constraints. Spin models
provide effective descriptions of a broad range of problems in
statistical mechanics, condensed matter physics, and lattice field
theory. In particular, they provide a natural setting in which to
investigate the effects of frustration, a phenomenon that
substantially modifies the structure of the low-energy configurations,
and may lead to unconventional ordering phenomena.

Frustration arises when local interactions cannot be simultaneously
minimized. In unfrustrated systems, the minimization requirements are
generally compatible, allowing the system to develop simple
lowest-energy states. Frustration instead introduces competing
constraints, so that minimizing the energies of parts of the
Hamiltonian does not simultaneously lead to the global minimization of
the energy. This situation can generate a large degeneracy of
low-energy configurations, which may somehow suppress conventional
long-range order. More generally, frustration may lead to the
emergence of competing phases, nontrivial correlations, topological
excitations, and unusual forms of criticality, see, e.g.,
Refs.~\cite{Ramirez-94, Moessner-01, Balents-10, Chalker-17, Diep-25}
for some review articles.

A common way to introduce frustration in spin systems is through
antiferromagnetic interactions on lattice geometries that inherently
give rise to frustration, such as triangular lattices.  However gauge
fields also provide a natural mechanism for generating frustration,
and complex spin systems coupled to a U(1) gauge field provide a
theoretical laboratory for studying the interplay between local gauge
invariance and frustration.  Indeed, to enforce U(1) gauge invariance
on the lattice, complex link variables $\lambda_{{\bm x},\mu}$ (with
$|\lambda_{{\bm x},\mu}|=1$) are introduced, mediating the interaction
between nearest-neighbor complex spins ${\bm z}_{\bm x}$ located at
the sites of the lattice, and for generic values of the variables
$\lambda_{{\bm x},\mu}$ frustration is typically present.

Several approaches can be identified when discussing gauge-induced
frustration. In the first approach the gauge field is fixed and
time-independent, thus constituting only a constant background for the
spin fluctuations. This is the case investigated, e.g., in \cite{TJ-83,
  HW-95, FT-95, FZ-01, APV-08}, where the two-dimensional XY model was
investigated in the presence of an external gauge field with a
constant and nontrivial magnetic flux. A different possibility is to
use a quenched random gauge field, which describes the case in which
the relaxation time scale of the gauge field is much larger than that
of the spin degrees of freedom, so it is effectively
out-of-equilibrium. This case corresponds to spin glass systems
\cite{MPV-book, Nishimori-book, Tahirietal-27}, and in this context by
far the most investigated instance has been that of a gauge field with
zero correlation length (i.e., gauge glass models with a random gauge
field~\cite{ON-93,Nishimori-01,AV-11}). For some studies using a gauge
field with nontrivial correlation length, see \cite{CP-19,
  Nishimori-24, Nishimori-25}, and \cite{BPV-24-z2Nv, BPV-24-unco} for
a different perspective.

In this paper we study instead the effects of dynamic thermalized
(annealed) frustrating gauge fields, considering the three-dimensional
(3D) compact lattice Abelian Higgs (CLAH) models with compact U(1)
gauge variables. The distribution of gauge variables $\lambda_{{\bm
    x},\mu}$ is in this case determined by the dynamics of the system,
and in particular by the interplay between the gauge self-coupling,
parametrized by $K$ (see below), and the gauge-scalar coupling
$J$. For $K\to\infty$ all link variables can be set to one (apart from
trivial gauge transformations), leading to standard ferromagnetic
interactions among the spins. With decreasing $K$, and in particular
for negative values of $K$, the bond variables can give rise to a
substantial frustration among the spin variables.  In particular, the
$K\to -\infty$ limit leads to full frustration, when the products of
the link variables along all plaquettes assume the value minus
one. Therefore, the phase diagram and phase transitions for negative
$K$ result from the competition of two opposing tendencies.  On the
one hand, the negative gauge coupling favors frustration. On the other
hand, as will be discussed in the following, the nearest-neighbor spin
interaction $-J\,{\rm Re}( \bar{\bm z}_{\bm x} \,\lambda_{{\bm
    x},\mu}\, {\bm z}_{{\bm x}+\hat\mu})$, with $J>0$, promotes
ferromagnetic alignment at large $J$ and favors positive plaquette
values, even when $K<0$.

CLAH models have been much investigated for positive gauge couplings
$K$, essentially because they provide lattice discretizations of the
scalar electrodynamics or Abelian Higgs field theories, which lead to
the realization of the Abelian Higgs mechanism, see, e.g.,
Refs.~\cite{BPV-25,OS-78,FS-79,FMS-81,CC-82,BF-83,FM-83,KK-85,KK-86,BN-86,
  BN-86-b,
  BN-87,MS-90,KKS-94,BFLLW-96,HT-96,KKLP-98,OT-98,CN-99,HS-00,KNS-02,
  SSSNH-02,SSNHS-03,MV-04,NSSS-04,SSS-04,WBJSS-05,CFIS-05,ODHIM-09,PV-19-CP,
  PV-19-AH3d,PV-20-largeNCP,BPV-20-hcAH,WB-21,BPV-21-gb,BPV-22,
  BPV-24-coH, BMV-26} for a (likely incomplete) list of references.
Their phase diagram crucially depends on the number of components of
the complex spins and the value of their integer charge.  Their phase
structure shows distinct regimes characterized by different behaviors
of gauge-invariant observables and correlation functions of the gauge
and spin variables.  In the space of couplings, there are ordered,
disordered and topological phases, crossover regimes, and phase
transitions whose nature is controlled by the interplay between matter
and gauge degrees of freedom. In particular, one also finds Higgs
transitions for multicharge scalar matter and sufficiently large
values of the gauge coupling~\cite{BPV-25}.

In this paper we extend the study of the phase diagram and phase
transitions of 3D CLAH models to negative gauge couplings $K$, when
the gauge interactions favor frustration among the complex spin
variables. We study the effects of the gauge frustration on the
collective behavior of the scalar matter, represented by one-component
or multicomponent complex spins, and the interplay of the strengths of
the matter-gauge and gauge couplings, which determines the main
features of the different phases and phase transitions.  As we shall
see, peculiar phases limited by topological ${\mathbb Z}_2$ and XY
gauge transition lines are found in both one-component and
multi-component CLAH models with negative gauge coupling.

The paper is organized as follows.  In Sec.~\ref{model} we introduce the CLAH
model and the observables considered in our study.  In Sec.~\ref{phadiadisc} we
discuss some general features of the phase diagram, and summarize the known
results for positive gauge coupling.  Sec.~\ref{numresN2} is devoted to the
presentation of our numerical results based on Monte Carlo (MC) simulations for
the phase diagram of the two-component unit-charge CLAH model with negative
gauge coupling, and the nature of the transitions along the lines separating
the different phases. In Sec.~\ref{numresN1} we present an analogous numerical
study for the one-component CLAH model. Finally, in Sec.~\ref{conclu} we
summarize and draw our conclusions.

\section{The lattice Abelian Higgs model with compact gauge variables}
\label{model}

\subsection{The model}
\label{mmm}

The 3D lattice Abelian Higgs model with compact gauge variables (CLAH
model) is a paradigmatic lattice gauge model describing charged
multicomponent scalar fields coupled to U(1) gauge variables.  The
compact formulation of U(1) gauge fields is achieved by considering
complex variables $\lambda_{{\bm x},\mu}\in$ U(1) associated with the
lattice links (starting from ${\bm x}$ toward the $\hat\mu$
direction). On a cubic lattice, the CLAH model with $N$-component
scalar fields ${\bm z}_{\bm x}$ of unit length ($\bar{\bm z}_{\bm
  x}\cdot {\bm z}_{\bm x}=1$), integer charge $q$, and maximal SU($N$)
global symmetry is defined by the partition function and lattice
Hamiltonian
\begin{eqnarray}
  Z = \sum_{\{{\bm z},\lambda\}} e^{-H},\qquad
  H = J N H_z + K H_g.
\label{hamco}
\end{eqnarray}
In this expression
\begin{eqnarray}
  H_z = - \sum_{{\bm x}, \mu} \left(
  \bar{\bm{z}}_{\bm x} \cdot \lambda_{{\bm x},\mu}^q\, {\bm z}_{{\bm
      x}+\hat\mu} + {\rm c.c.}\right)
\label{HamZ}
\end{eqnarray}
is the interaction term for the scalar fields of charge $q$, where the
sum is over all lattice links of the cubic lattice, and
\begin{eqnarray}
  H_g &=& - \sum_{{\bm x},\,\mu<\nu} \left(\Pi_{{\bm
      x},\mu\nu}+{\rm c.c.}\right),\label{Hamg} \\
\Pi_{{\bm x},\mu\nu} &=&
    \lambda_{{\bm x},{\mu}} \,\lambda_{{\bm x}+\hat{\mu},{\nu}}
    \,\bar{\lambda}_{{\bm x}+\hat{\nu},{\mu}} \,\bar{\lambda}_{{\bm
        x},{\nu}},
\label{plaquette}
\end{eqnarray}
where the sum is over the lattice plaquettes. Note that within compact
formulations of the gauge variables the charge $q$ of the matter field
cannot be eliminated by a redefinition of the fields, thus $q$
represents a further Hamiltonian parameter of the model.  In our
numerical study we consider cubic lattices of linear size $L$ with
periodic boundary conditions.

It is immediate to see that the Hamiltonian in Eq.~\eqref{hamco} is
invariant under the U(1) gauge transformation
\begin{eqnarray}
  \lambda_{{\bm x},\mu} \to U_{\bm x} \lambda_{{\bm x},\mu} \bar{U}_{{\bm x}+
    \hat{\mu}},\qquad  {\bm z}_{\bm x} \to  U_{\bm x}^q {\bm z}_{\bm x},
  \label{gautra}
\end{eqnarray}
with $U_{\bm x}$ a position dependent U(1) factor.  It is equally
simple to see that the Hamiltonian is invariant under the U($N$)
global transformation ${\bm z}_{\bm x} \to V {\bm z}_{\bm x}$, where
$V$ is a constant (i.e., position independent) $N\times N$ unitary
matrix, and the global symmetry of the system is thus SU($N$) or, more
precisely, SU($N$)$/\mathbb{Z}_N$, where $\mathbb{Z}_N\subset U(1)$ is the
center of SU($N$).

Since the staggered spin change of variable 
\begin{equation}
{\bm z}_{\bm x} \to  (-1)^{\sum_{\mu}x_{\mu}} {\bm z}_{\bm x}
\end{equation}
in the partition function effectively corresponds to changing the sign
of $J$ in the Hamiltonian, the partition function is an even function
of $J$.  It is also easy to check that, for even values of the charge
$q$, the partition function is also invariant under $K\to -K$. Indeed,
one can effectively change the sign of the gauge coupling $K$ by
performing a link transformation limited to the ${\mathbb Z}_2$
subgroup, such as
\begin{eqnarray}
\lambda_{{\bm x},\mu} \to
(-1)^{\sum_{\nu<\mu}x_\nu} 
\lambda_{{\bm x},\mu}.
\label{kmktra}
\end{eqnarray}
Since this gauge-variable change does not affect the spin Hamiltonian
term $H_z$ for even $q$, the partition function is in this case an
even function of $K$.  From these properties it follows that to study
the phase diagram of the CLAH models we can always restrict to
consider $J\ge 0$ and, for even values of $q$, also $K\ge 0$.  The
situation drastically changes for odd values of $q$, where a negative
value of the gauge coupling induces frustrated interactions among the
spin variables.

In the simplest one-component ($N=1$) case, the Hamiltonian
(\ref{HamZ}) can be simplified by choosing the unitary gauge, obtained
by fixing $z_{\bm x}=1$. Therefore the $N=1$ CLAH Hamiltonian can be
equivalently written as
\begin{eqnarray}
H_{N=1} = - 2 J \sum_{{\bm x}, \mu} {\rm Re}\,\lambda_{{\bm x},\mu}^q
- 2 K \sum_{{\bm x},\,\mu<\nu} {\rm Re}\,\Pi_{{\bm
      x},\mu\nu},
\label{HamN1}
\end{eqnarray}
which can be seen as a compact U(1) gauge model with a linear
gauge-symmetry breaking controlled by the coupling $J$, see
\cite{BPV-21-gb}.

\subsection{Observables}
\label{obs}

To characterize the different phases, and phase transitions,
associated with the breaking of the SU($N$) global symmetry, we
consider correlations of the gauge-invariant Hermitean bilinear
operator
\begin{equation}
Q^{ab}_{\bm x} = \bar{z}_{\bm x}^a z_{\bm x}^b - {1\over N} \delta^{ab}.
\label{qdef}
\end{equation}
Its two-point correlation function is defined as
\begin{equation}
G_Q({\bm x}-{\bm y}) = \langle {\rm Tr}\, Q_{\bm x} Q_{\bm y}\, \rangle,  
\label{gxyp}
\end{equation}
where the translation invariance of the system with periodic boundary
conditions is taken into account.  The corresponding susceptibility
and second-moment correlation length are defined as
\begin{eqnarray}
  \chi_Q &=&\sum_{\bm x} G_Q({\bm x}),\label{chidef}\\
  \xi^2 &=&  {1\over 4 \sin^2 (\pi/L)}
{\widetilde{G}_Q({\bm 0}) - \widetilde{G}_Q({\bm p}_m)\over 
\widetilde{G}_Q({\bm p}_m)},
\label{xidefpb}
\end{eqnarray}
where $\widetilde{G}_Q({\bm p})=\sum_{{\bm x}} e^{i{\bm p}\cdot {\bm
    x}} G_Q({\bm x})$ is the Fourier transform of $G_Q({\bm x})$, and
${\bm p}_m = (2\pi/L,0,0)$ is the minimum nonzero lattice momentum (we
always use symmetric lattices with periodic boundary conditions).  To
investigate the critical behaviors associated with the spontaneous
symmetry breaking of the global symmetry, it is convenient to study
renormalization-group invariant quantities, such as the ratio
\begin{equation}
R_\xi = \xi/L,
\label{rxi}
\end{equation}
and the Binder parameter
\begin{equation}
U = {\langle \mu_2^2\rangle \over \langle \mu_2 \rangle^2} \,, \qquad
\mu_2 = 
\sum_{{\bm x},{\bm y}} {\rm Tr}\,Q_{\bm x} Q_{\bm y}.
\label{binderdef}
\end{equation}

To study topological phase transitions without local order parameters,
which are not associated with the spontaneous breaking of any global
symmetry, one may focus on quantities related to the energy
density. We thus study the energy density and the specific heat
\begin{eqnarray}
  E &=&  \frac{1}{V} \langle H \rangle,\qquad V=L^3,
  \label{enedef}\\
  C &=&  \frac{1}{V}
  \langle(H-\langle H\rangle)^2\rangle,
  \label{cvdef}
\end{eqnarray}
and, to reduce the contribution of analytic background terms
\cite{SSNHS-03, BPV-24-coH}, in some cases we also consider the third
and fourth cumulants
\begin{eqnarray}
  H_3 &=&  \frac{1}{V} \left\langle (H-\langle H\rangle)^3 \right\rangle, \label{H3}\\
  H_4 &=&  \frac{1}{V}\left[\left\langle (H-\langle H\rangle)^4 \right\rangle 
-3\left\langle (H-\langle H\rangle)^2\right\rangle^2\right]. \label{H4}
\end{eqnarray}

Finally, to monitor the amount of frustration in the system we consider the 
average plaquette defined as
\begin{eqnarray}
  W = \frac{1}{3V} \sum_{{\bm x},\mu<\nu}
    \langle {\rm Re}\;\Pi_{{\bm x},\mu\nu}\rangle.
  \label{wplaque}
\end{eqnarray}

\section{The phase diagrams of CLAH models}
\label{phadiadisc}

\subsection{Some general facts}
\label{phadiagen}

We now summarize some general features of the phase diagram of CLAH
models, which can be useful as guiding principles in numerical
investigations.

For $K\to\infty$ the configurations which dominate the partition
function are those which minimize $H_g$, hence those characterized by
having all the plaquettes $\Pi_{{\bm x},\mu\nu}$ equal to
one. Neglecting finite volume effects, these configurations are gauge
equivalent to the trivial configuration $\lambda_{{\bm x},\mu}=1$. In
the limit $K\to\infty$, CLAH models with SU($N$) symmetry thus reduce
to real O($M$) vector models with $M=2N$, which display continuous
phase transitions at a finite value of $J$. Note, however, that the
fixed points associated with these continuous transitions can be shown
to be unstable with respect to the introduction of a gauge coupling
\cite{PV-19-AH3d}, thus continuous transitions in the O($2N$)
universality class are not expected for large but finite values of
$K$.

In the opposite limit $K\to -\infty$, the relevant configurations are
those which maximize $H_g$, giving rise to fully-frustrated plaquettes
$\Pi_{{\bm x},\mu\nu} \to -1$.  In this limit, using gauge invariance,
one can fix the gauge variables to the frozen gauge configuration
\begin{eqnarray}
  \lambda_{{\bm x},1}^{\rm (ff)}= 1,\quad
  \lambda_{{\bm x},2}^{\rm (ff)}= (-1)^{x_1},\quad
  \lambda_{{\bm x},3}^{\rm (ff)}= (-1)^{x_1+x_2},
  \label{fruconf}
\end{eqnarray}
so that for all plaquettes $\Pi_{{\bm x},\mu\nu} = -1$ (note that this
choice of link values is consistent with the periodic boundary
conditions of finite systems for even size $L$). Once gauge fields are
fixed in this way, for odd values of the charge $q$ the CLAH model
reduces to the fully frustrated O($M$) vector model with $M=2N$,
characterized by the Hamiltonian
\begin{equation}
H_{{\rm O}(M)} = - J M \sum_{{\bm x}, \mu} \lambda_{{\bm x},\mu}^{\rm (ff)}\,
{\bm s}_{\bm x} \cdot \, {\bm s}_{{\bm x}+\hat\mu},
  \label{homfru}
\end{equation}
where ${\bm s}_{\bm x}$ is an $M$-component unit-length real vector.

In the limit $J\to\infty$ it is instead the term $H_z$ of the action
which has to be minimized, hence the relevant configurations are those
satisfying the constraint ${\bm z}_{\bm x}=\lambda^q_{{\bm x,\mu}}{\bm
  z}_{{\bm x}+\hat{\mu}}$.  By using this constraint iteratively on
the sites of a plaquette we then get $\Pi_{{\bm x},\mu\nu}^q =
1$. Thus, the gauge variables can be chosen (once again up to gauge
transformations) to be elements of $\mathbb{Z}_q$ in the large $J$
limit. In the basic case $q=1$ that will be studied in the following,
this implies that for any fixed $K$ value there is no frustration when
$J\to\infty$.

CLAH models with $N=1$ and $N>1$ are expected to have quite different
phase diagrams. Models with $N>1$ are indeed characterized by the
presence of the global SU($N$) symmetry, whose spontaneous breaking
driven by the condensation of the gauge invariant bilinear $Q_{\bm
  x}$, cf. Eq.~\eqref{qdef}, gives rise to spin ordered phases. For
$N=1$ no global symmetry is present, and the phase diagram is simpler.
In particular, using a convergent cluster expansion, for the $N=1$
model with unit charge ($q=1$) it is possible to show that the free
energy density is analytic in a large region of the phase diagram,
which starts close to $K=0$ axis and is bounded, for large $|K|$, by
the curve $|K/J|<$~const \cite{OS-78, FS-79}. Note that the original
argument was developed for $K>0$, but being based on a convergent
series expansion in $K$ it also holds for negative $K$ values.

Despite these differences, the ordered phase of the $N>1$ models
presents similarities with the $N=1$ case. Let us consider for the
sake of simplicity the $N=2$ models, but the argument can be
easily extended to generic $N>1$. We can fix the unitary gauge by
imposing the spins ${\bm z}_{\bm x}$ to be of the form
\begin{equation}
  {\bm z}_{\bm x}=\left(\begin{array}{cc} a_{\bm x} \\
    w_{\bm x} \end{array}\right),
\qquad a_{\bm x}^2+|w_{\bm x}|^2=1,
\end{equation}
with $a_{\bm x}\in\mathbb{R}$, $a_{\bm x}\ge 0$ and $w_{\bm
  x}\in\mathbb{C}$.  The tensor order parameter $Q_{\bm x}$
is then given by
\begin{equation}
  Q_{\bm x}=\left(\begin{array}{cc} a_{\bm x}^2-\tfrac{1}{2} &
    \bar{w}_{\bm x}a_{\bm x}\\
a_{\bm x}w_{\bm x} & |w_{\bm x}|^2-\tfrac{1}{2}\end{array}\right).
\end{equation}
If we assume the ordered phase to be translation invariant, then
$a_{\bm x}$ and $w_{\bm x}$ must be independent of the position ${\bm
  x}$.  Neglecting fluctuations we can thus write
\begin{equation}
{\bm z}_{\bm x}=\left(\begin{array}{cc} a \\ w \end{array}\right),
\end{equation}
and the term $H_z$ of the Hamiltonian takes the form
\begin{eqnarray}
- 2JN\sum_{{\bm x}, \mu} {\rm Re}\,\lambda_{{\bm x},\mu}^q\, 
  \bar{\bm{z}}_{\bm x} \cdot {\bm z}_{{\bm
      x}+\hat\mu}  =
  - 2JN\sum_{{\bm x}, \mu} {\rm Re}\,\lambda_{{\bm x},\mu}^q.
\label{ham_largeJ}
\end{eqnarray}
By neglecting fluctuations of $Q_{\bm x}$ we thus recover the
Hamiltonian of the $N=1$ model in the unitary gauge (but for the
substitution $J\to JN$), cf. Eq.~\eqref{HamN1}.  This means that in
the ordered phase of the $N>1$ models the gauge-invariant dynamics of
the gauge fields should be the same as that of the $N=1$ model with
the same charge.

\subsection{Phase diagrams for $K\ge 0$}
\label{phadiakge0}

Several results have been already obtained for the phase diagrams and
transition lines of 3D CLAH models for nonnegative gauge couplings,
$K\ge 0$, see, e.g.,
Refs.~\cite{OS-78,FS-79,FMS-81,CC-82,BF-83,FM-83,KK-85,KK-86,BN-86,BN-86-b,
  BN-87,MS-90,KKS-94,BFLLW-96,HT-96,KKLP-98,OT-98,CN-99,HS-00,KNS-02,
  SSSNH-02,SSNHS-03,MV-04,NSSS-04,SSS-04,WBJSS-05,CFIS-05,ODHIM-09,PV-19-CP,
  PV-19-AH3d,PV-20-largeNCP,BPV-20-hcAH,BPV-21-gb,WB-21,BPV-22, BPV-24-coH, BMV-26},
and in particular Ref.~\cite{BPV-25} for a review of
results. Here, we briefly summarize the main features depending on the
number $N$ of components, in particular whether $N=1$ or $N\ge 2$, and
the charge $q$, in particular whether $q=1$ or $q\ge 2$.

\begin{figure}[tbp]
  \includegraphics*[width=0.90\columnwidth]{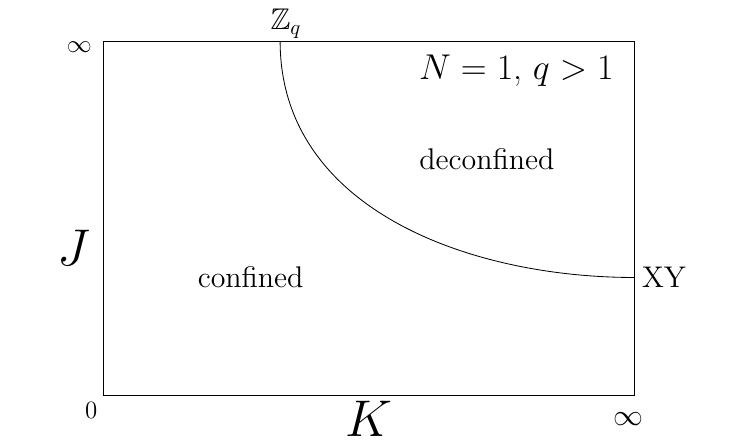}
  \caption{Sketch of the phase diagram of the 3D one-component CLAH
    models with positive gauge coupling $K\ge 0$, for charges $q$
    larger than one. A confined phase (for small $J$ or small $K$) and
    a deconfined phase (for large $J$ and $K$) are present, separated
    by a deconfinement transition line.  For $K\to\infty$ and
    $J\to\infty$ the model reduces to the XY vector model and to the
    lattice ${\mathbb Z}_q$ gauge model with Wilson action,
    respectively.
    \label{phadiaN1}}
\end{figure}

For $N=1$ and $q=1$ the phase diagram is trivial, presenting only one
phase, without phase transitions at any finite $J,K\ge 0$.  There is
only an isolated critical point for $K\to \infty$, where the model
becomes equivalent to the lattice O(2) vector or XY model.  Two
phases are instead present for $N=1$ and $q>1$, as shown in
Fig.~\ref{phadiaN1}, separated by a topological transition line
belonging to the 3D ${\mathbb Z}_q$ gauge universality class, starting
from the XY point at $K=\infty$ and ending on the $J=\infty$ line.

\begin{figure}[tbp]
  \includegraphics*[width=0.90\columnwidth]{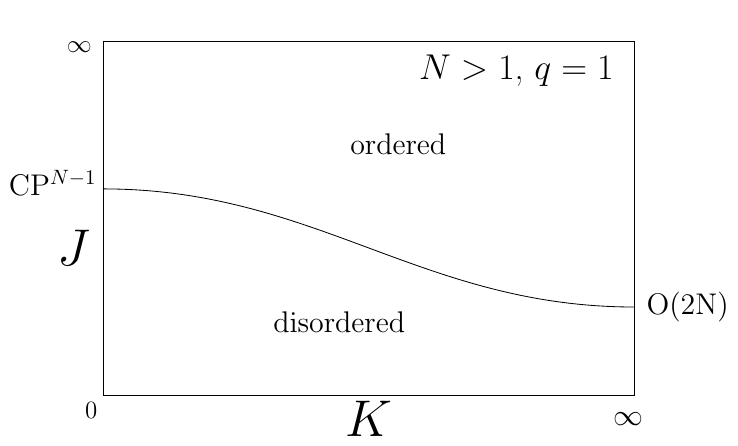}
  \caption{Sketch of the phase diagram of the 3D multicomponent ($N>1$) CLAH
    models with positive gauge coupling $K\ge 0$, for scalar matter
    with charge $q=1$. There are two phases, disordered and ordered,
    distinguished by the condensation of the bilinear operator $Q_{\bm x}$
    defined in Eq.~(\ref{qdef}).  The CLAH models become equivalent to
    the CP$^{N-1}$ model for $K=0$, and to the O($2N$) vector model
    for $K\to\infty$ (apart from trivial gauge transformations).
    \label{phadiaN2q1}}
\end{figure}

\begin{figure}[tbp]
  \includegraphics*[width=0.90\columnwidth]{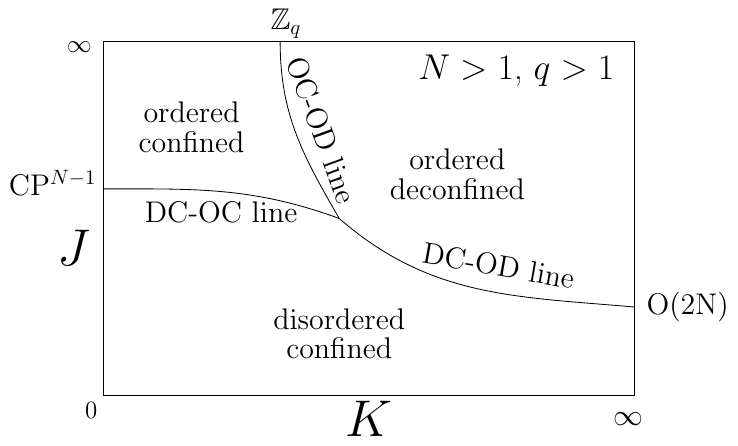}
  \caption{Sketch of the phase diagrams of the 3D multicomponent
    ($N>1$) CLAH models with positive gauge coupling $K\ge 0$, for
    multicharge scalar matter ($q>1$).  These systems present three phases:
    the disordered-confined (DC), the ordered-deconfined (OD), and the
    ordered-confined (OC) phase. They are separated by different
    DC-OD, DC-OC and OC-OD transition lines~\cite{BPV-25}.  In
    particular, continuous transitions are found along the DC-OD line
    for sufficiently large $N$~\cite{BMV-26}, which belong to the 3D
    Abelian Higgs universality class~\cite{BPV-25,BPV-20-hcAH,BMV-26}.
    They become equivalent to the CP$^{N-1}$ model for $K=0$, and to
    the O($2N$) vector model for $K\to\infty$ (apart from trivial
    gauge transformations). Moreover, in the limit $J\to\infty$ they
    become equivalent to ${\mathbb Z}_q$ gauge models.
    \label{phadiaN2q2}}
\end{figure}

Multicomponent ($N>1$) CLAH models present more complex phase diagrams,
with notable differences between the cases of single-charge ($q=1$)
and multi-charge ($q\ge 2$) spin variables, as sketched in
Figs.~\ref{phadiaN2q1} and \ref{phadiaN2q2}.

In Fig.~\ref{phadiaN2q1} we show the phase diagrams of single-charge
$q=1$ multicomponent ($N>1$) CLAH models, which display two phases: a
disordered phase for small $J$ and an ordered phase for large $J$.
For $K=0$ the model is equivalent to a particular lattice formulation
of the CP$^{N-1}$ model, which undergoes a phase transition at a
finite value of $J$; this phase transition is continuous for $N=2$,
belonging to the Heisenberg universality class, and first order for
any $N\ge 3$~\cite{PV-19-CP,PV-20-largeNCP}.  In the $K\to\infty$
limit the model reduces to an O($2N$) vector model, which presents a
transition at a finite value of $J$, as well.  The disordered and
ordered phases are separated by a single transition line,
characterized by the breaking of the SU($N$) symmetry, where the
gauge-invariant bilinear operator (\ref{qdef}) plays the role of order
parameter.  The phase transitions are expected to be analogous to those
of the 3D CP$^{N-1}$ model for $K=0$, as also confirmed by numerical
analyses~\cite{PV-19-AH3d}.

As sketched in Fig.~\ref{phadiaN2q2}, multicharged ($q>1$) and
multicomponent ($N\ge 2$) CLAH models show a more complex phase
diagram~\cite{BPV-25}.  Indeed, their phase diagrams show a disordered
(confined) phase for small values of $J$ and two ordered phases for
large values of $J$. The transitions between the disordered and the
ordered phases are associated with the breaking of the global SU($N$)
symmetry, while the transition between the two ordered phase is a
deconfinement transition.  These models are particularly interesting
because the transitions between the ordered-deconfined and
disordered-confined phases are continuous for $N\ge N^*$ with
$N^*=9(1)$~\cite{BMV-26}, and belong to the 3D Abelian Higgs
universality class, thus realizing the continuum limit of the 3D
Abelian Higgs field theory or scalar
electrodynamics~\cite{BPV-25}.

\subsection{Negative gauge couplings}
\label{negcoup}

In Sec.~\ref{phadiagen} we have shown that the free energy density of
CLAH models is an even function of $K$ when $q$ is an even number, so
the phase diagram is completely determined in these cases. Little is
instead known about the phase diagrams of CLAH models with odd charge
$q$ for negative values of the gauge coupling $K$.

For odd $q$ a negative gauge coupling induces a nontrivial frustration
on the complex spin variables, which is expected to have notable
effects on the thermodynamic behavior of the system.  We recall that
frustrated spin interactions may arise when the product of the bonds
surrounding any plaquette is negative, i.e., when one cannot minimize
all the spin Hamiltonian terms around a plaquette.  Indeed, a negative
value of the gauge coupling $K$ favors negative values of the average
plaquette $W$, cf. Eq.~(\ref{wplaque}), inducing frustrated
interactions in the spin Hamiltonian term $H_z$, cf. Eq.~(\ref{HamZ}).

In particular, in Sec.~\ref{phadiagen} we have shown that CLAH models
reduce to fully-frustrated O($M$) vector models in the $K\to-\infty$
limit, with $M=2N$.  The critical properties of these models have
however not been so far thoroughly investigated for generic values of
$M$.  Some numerical analyses have been carried out for the
fully-frustrated O($M$) vector models with $M\le 3$, providing
evidence of first-order transitions for both $M=2$ and $3$, see, e.g.,
Refs.~\cite{NHD-10,NHD-11}.  On the other hand, a continuous
transition is found for the fully-frustrated spherical model
limit~\cite{DPTV-80}, which should correspond to the limit
$M\to\infty$, at the critical point $J_c\approx 0.89062$ (obtained
using the results of Ref.~\cite{DPTV-80}). This may suggest the
existence of an intermediate finite value $M^*>3$, such that the
transitions are first order for $M<M^*$ and continuous for $M>M^*$
(actually, one cannot exclude that such finite $M^*$ does not exist,
and the continuous transition of the spherical model is only realized
at $M=\infty$ strictly).

It is therefore natural to expect that the order-disorder transition
line (for $q=1$) or the DC-OC transition line (for $q>1$, see
Fig.~\ref{phadiaN2q2}) emerging from $K=0$ extends into the $K<0$
region, eventually reaching the order-disorder transition of the fully
frustrated model at $K=-\infty$.

For $N=2$, the transitions along this line belong to the O(3)
universality class for $K\gtrsim 0$, and it is reasonable to expect
the same behavior for sufficiently small negative values of
$K$. However, no results are currently available for the fully
frustrated O(4) model (corresponding to $N=2$), corresponding to $K\to
-\infty$ limit.  If the order-disorder transition of this model is
discontinuous (as is the case for $M=2,3$), the first-order nature of
the transition is expected to persist for sufficiently large negative
values of $K$. The O(3) transition line must therefore change its
nature at some finite negative value of $K$, becoming a first-order
transition line.  If, instead, the transition of the fully frustrated
model is continuous, several scenarios are possible. The transition
line could remain in the O(3) universality class for all finite
negative values of $K$, with the fully frustrated model governing only
the transition at $K=-\infty$ (analogously to what happens at
$K=+\infty$ with the O($2N$) model). Alternatively, the transition
line could change its universality class at some finite negative value
of $K$, after meeting a multicritical point. Finally, an intermediate
region of first-order transitions could emerge.

So far, we have implicitly assumed a certain continuity between the
ordered phases at $K=0$ and $K=-\infty$. However, we cannot exclude
the possibility of additional phase transitions arising from the
competition between the $H_z$ and $H_g$ terms in the Hamiltonian.  For
single-charge models, the limit $J\to\infty$ is expected to be trivial
for any finite $K$, including negative values, since in this limit the
link variables are forced to take the trivial value $\lambda_{{\bm
    x},\mu}=1$, up to gauge transformations (see
Sec.~\ref{phadiagen}). Therefore, no phase transitions are expected
along the $J=\infty$ line at finite $K$.  Using the relation between
the gauge-field dynamics in the ordered phase of the $N>1$ models and
those of the $N=1$ model, discussed in Sec.~\ref{phadiagen}, we can
draw a stronger conclusion. Since the $N>1$ models are in the ordered
phase for sufficiently large $J$, and since the absence of phase
transitions at sufficiently large but finite $J$ can be rigorously
established for the $N=1$ single-charge model at any finite
$K$~\cite{OS-78,FS-79}, we expect the same result to hold for the
$N>1$ single-charge models.  Consequently, if an additional phase
exists in the phase diagram of single-charge CLAH models, the
transition lines separating it from the other phases cannot reach the
boundary at $J=+\infty$ of the phase diagram.

\begin{figure}[tbp]
  \includegraphics*[width=0.90\columnwidth]{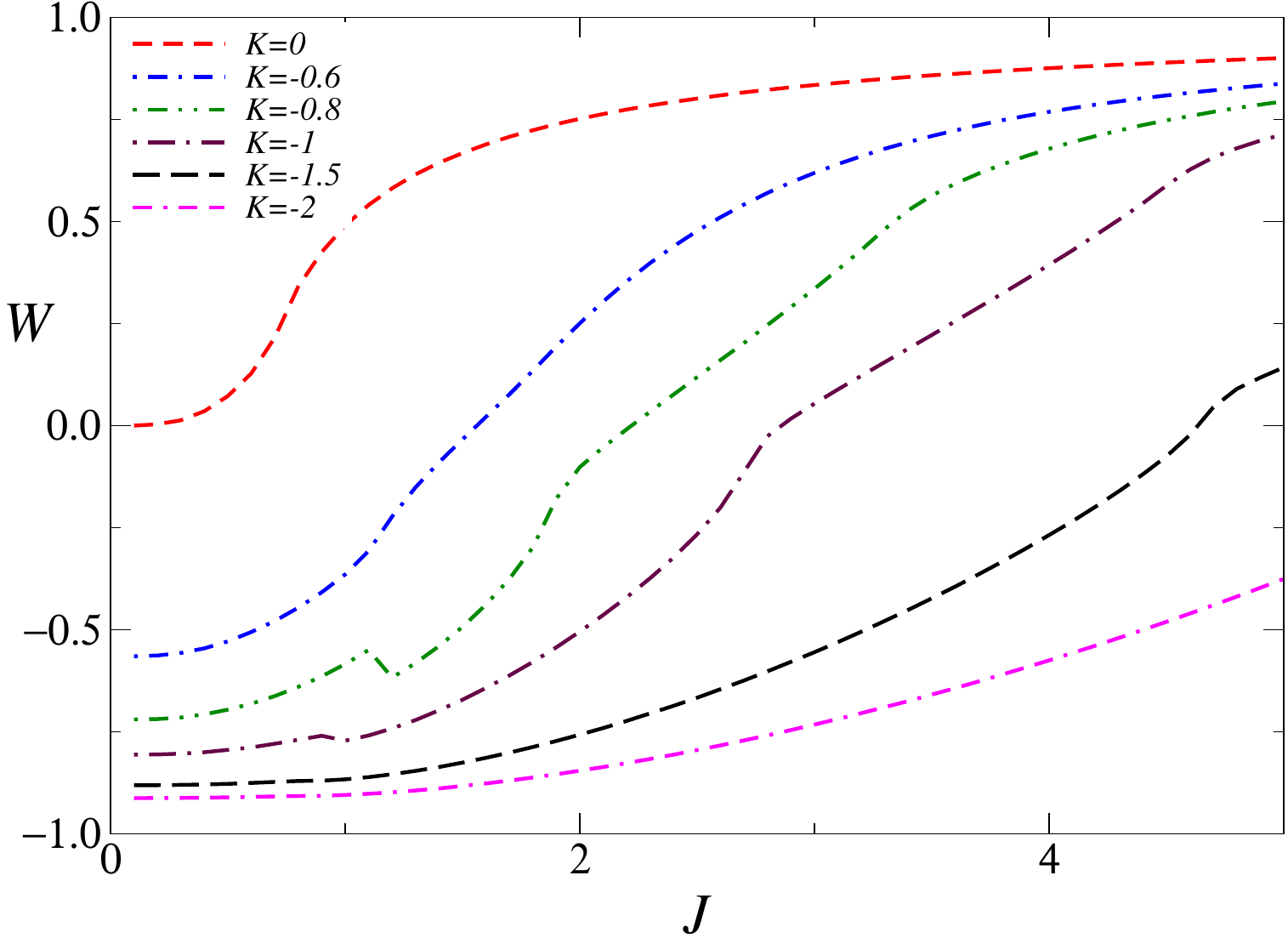} 
  \includegraphics*[width=0.90\columnwidth]{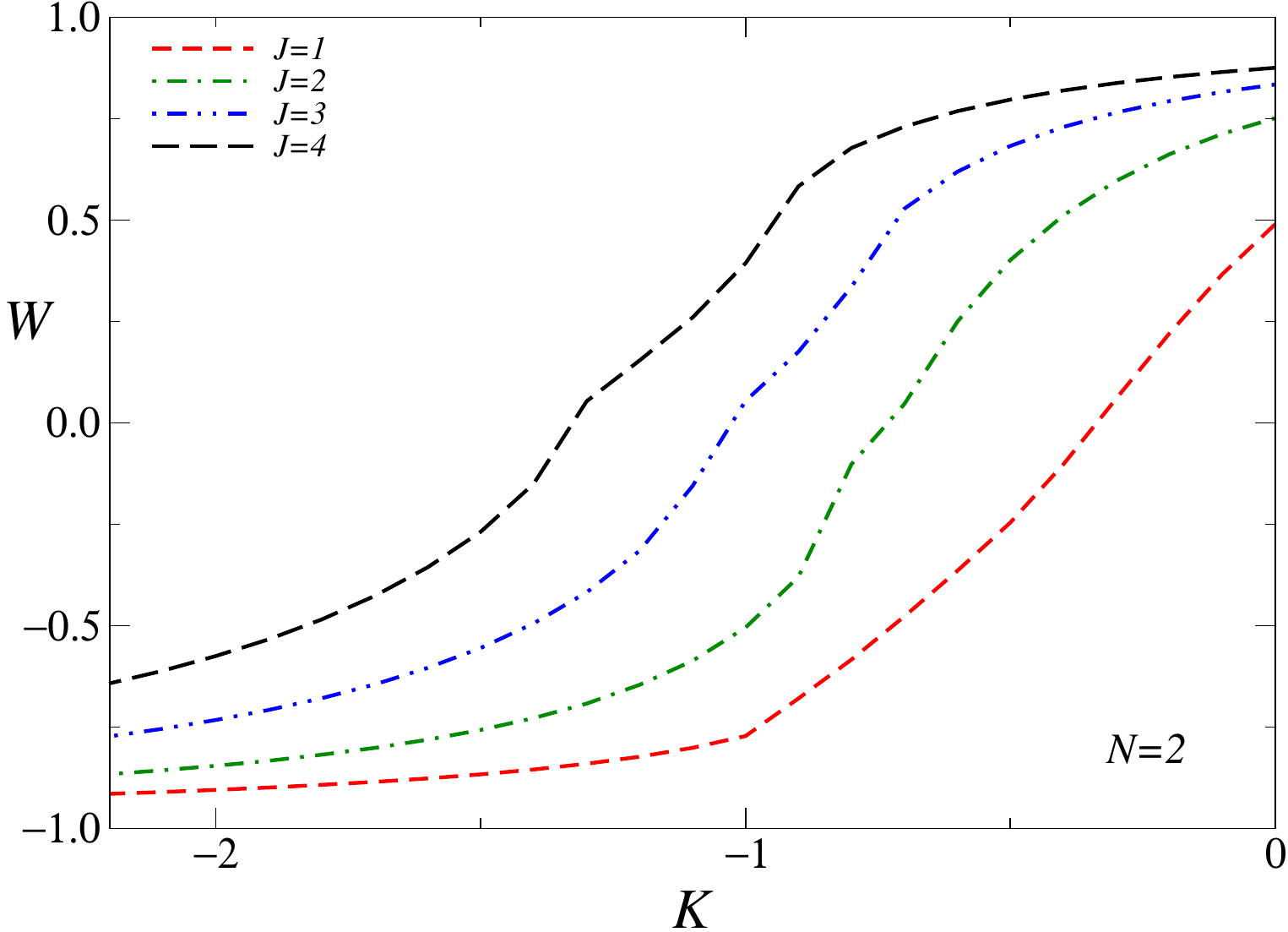}
  \caption{Behavior of the plaquette $W$ defined in
    Eq.~\eqref{wplaque} in the negative gauge coupling $K$ region of
    the $N=2$ single charged model.  These data have been obtained by
    performing scans using a relatively small lattice size, $L=10$,
    keeping $K$ fixed and changing $J$ (top panel) or keeping $J$
    fixed and changing $K$ (bottom panel). Errors are typically
    smaller than the thickness of the lines.
    \label{Wfigs}}
\end{figure}

That something nontrivial happens for moderately large values of $J$
and $K$ can be already guessed from Fig.~\ref{Wfigs}, where the
average values of the plaquette $W$, see Eq.~\eqref{wplaque}, are
reported as a function of $J$ and $K$ for the single-charge $N=2$
model in the $K\le 0$ region.  Note that for any fixed value of $K$
there is no significant frustration if $J$ is large enough.  A
detailed analysis of the phase diagram of this model will be the
subject of the next section.

As already mentioned, for higher even charges, the phase diagram for
$K<0$ is analogous to that $K>0$, due to the transformation
(\ref{kmktra}), which brings $K\to -K$ without changing the spin
Hamiltonian term (\ref{HamZ}).  Therefore, there is a ${\mathbb Z}_q$
gauge transition for $J\to\infty$ even for $K<0$, at $K= - K_c$. The
situation is less clear for multicharge models with odd $q$.  In the
following we mostly discuss the negative-$K$ phase diagram of CLAH
models in the most interesting single-charge case.  Therefore, unless
specified, in the following CLAH models are understood to be those
with $q=1$.

\section{Numerical results for the two-component CLAH model}
\label{numresN2}

In this section the negative-$K$ phase diagram and phase transitions
of the $N=2$ single-charge ($q=1$) CLAH model (\ref{hamco}) are
investigated numerically.  Like the case of positive gauge couplings,
see Fig.~\ref{phadiaN2q1}, we generally expect the existence of
phases associated with the condensation of the gauge-invariant
bilinear spin operator $Q_{\bm x}^{ab}$, cf. Eq.~(\ref{qdef}), leading
to disordered and ordered phases in low- and large-$J$ regions,
respectively.

\subsection{Monte Carlo simulations}
\label{mcsim}

The MC simulations are based on local overrelaxed upgradings of the spin and
gauge variables, exploiting the linearity of the single-charge CLAH Hamiltonian
with respect to each lattice variable. More precisely, the local upgrading of
each variable consists in a mixing of microcanonical~\cite{Adler-81,
Creutz:1987xi} (keeping the energy unchanged) and standard Metropolis
updates~\cite{MRRTT-53} of the lattice variables, to ensure
ergodicity~\cite{PV-19-AH3d}. One global update step consists of a number of
sweeps of all lattice variables: one sweep of Metropolis upgradings (with an
average acceptance of 30\% at least) and ten sweeps of microcanonical updates.
For each simulated point we gathered statistics (after thermalization) of
about O($10^7$) update steps. 

Cubic lattices of linear size $L$ with periodic boundary conditions
are always considered.  To determine the location and nature of the
transitions between the different phases of the $N=2$ CLAH model, we
analyze the finite-size behavior of results obtained by MC simulations
for various values of $K<0$, looking for their asymptotic finite-size
scaling (FSS) behaviors, which may distinguish the continuous from the
first-order transitions (see, e.g., Refs.~\cite{BMV-26,PV-24} for a
thorough discussion of the main different features of the FSS
behaviors at the two types of transitions).

\subsection{General features of the phase diagram}

Before discussing the transition lines appearing in the phase diagram
of the $N=2$ single-charge model in detail, it is useful to first
obtain an overall picture of the role played by frustration.  To this
end, in Fig.~\ref{Wfigs} we report the results for the average
plaquette $W$, cf. Eq.~\eqref{wplaque}, obtained from several scans
performed at fixed $K$ or $J$. Although these scans were carried out
at a single lattice size, several distinct regimes in the behavior of
$W$ can be identified, separated by points at which $W$ appear to
exhibit nonsmooth behaviors.

It is reasonable to expect the order-disorder transition line to lie
at $J\lesssim 1$. Indeed, for $K=0$, the O(3) transition of the CP$^1$
model is located at $J=0.7102(1)$, see Ref.~\cite{PV-19-CP}, while for
$K\to -\infty$, the large-$N$ limit predicts $J_c\approx 0.89062$
(using the results of Ref.~\cite{DPTV-80}) and for $M=2$
Ref.~\cite{NHD-10} reports $J_c\approx 0.735$. However,
Fig.~\ref{Wfigs} suggests that this may not be the only transition
line present in the phase diagram.

In fact, even for values of $J$ significantly larger than 1, two
anomalous behaviors are encountered when scanning in $K$, as
illustrated, for example, by the results for $J=3$ and $J=4$ in the
bottom panel of Fig.~\ref{Wfigs}, which may hint at the existence of
further high-$J$ transition lines.  One of them occurs (at least for
the values of $J$ and $K$ explored) when $W\approx 0$ and likely
delimits a phase connected to the fully frustrated model at
$K=-\infty$, characterized by a significant degree of frustration. The
other transition line, instead, is found at distinctly positive values
of the plaquette ($W\approx 0.5$), making it unlikely that this
transition is associated with frustration. A more natural
interpretation is that it arises from the competition between the
$H_z$ and $H_g$ interaction terms in the Hamiltonian.  Note that both
transition lines appear to shift toward larger values of $J$ as $K$
becomes increasingly negative, consistently with the previous
analysis, which excludes the presence of transitions at fixed finite
$K$ and sufficiently large $J$.

It is important to emphasize that, at this stage, these
interpretations should be regarded only as working hypotheses. They
must be tested through careful FSS analyses to establish whether
genuine phase transitions are present and, if so, to determine their
universality classes.  This is the purpose of the analysis presented
in the remainder of this section.

\subsection{Continuous O(3) transitions extending to $K<0$}
\label{conttra}

We now present results obtained along the extension of the O(3)
continuous transition line crossing the $K=0$ axis at $J_c=0.7102(1)$
(see Ref.~\cite{PV-19-CP}), showing that this line continues up to a
finite negative value $K^*$ of $K$, with $K^*\lesssim -0.7$. For this
purpose we report some results of MC simulations along the lines
$K=-0.5$ and $K=-0.7$, obtained using lattices of linear size up to
$L=32$.

\begin{figure}[tbp]
  \includegraphics*[width=0.90\columnwidth]{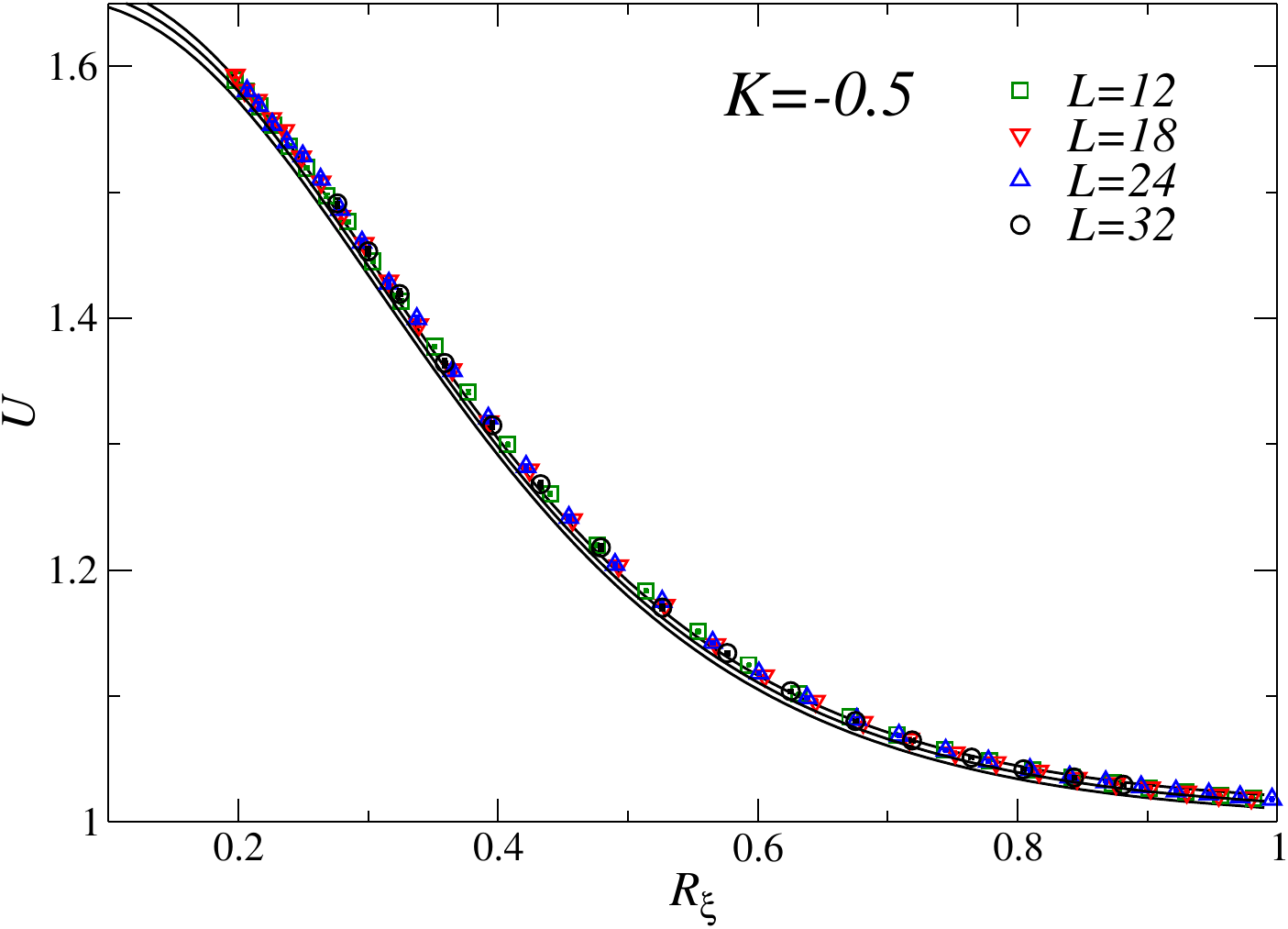} 
  \includegraphics*[width=0.90\columnwidth]{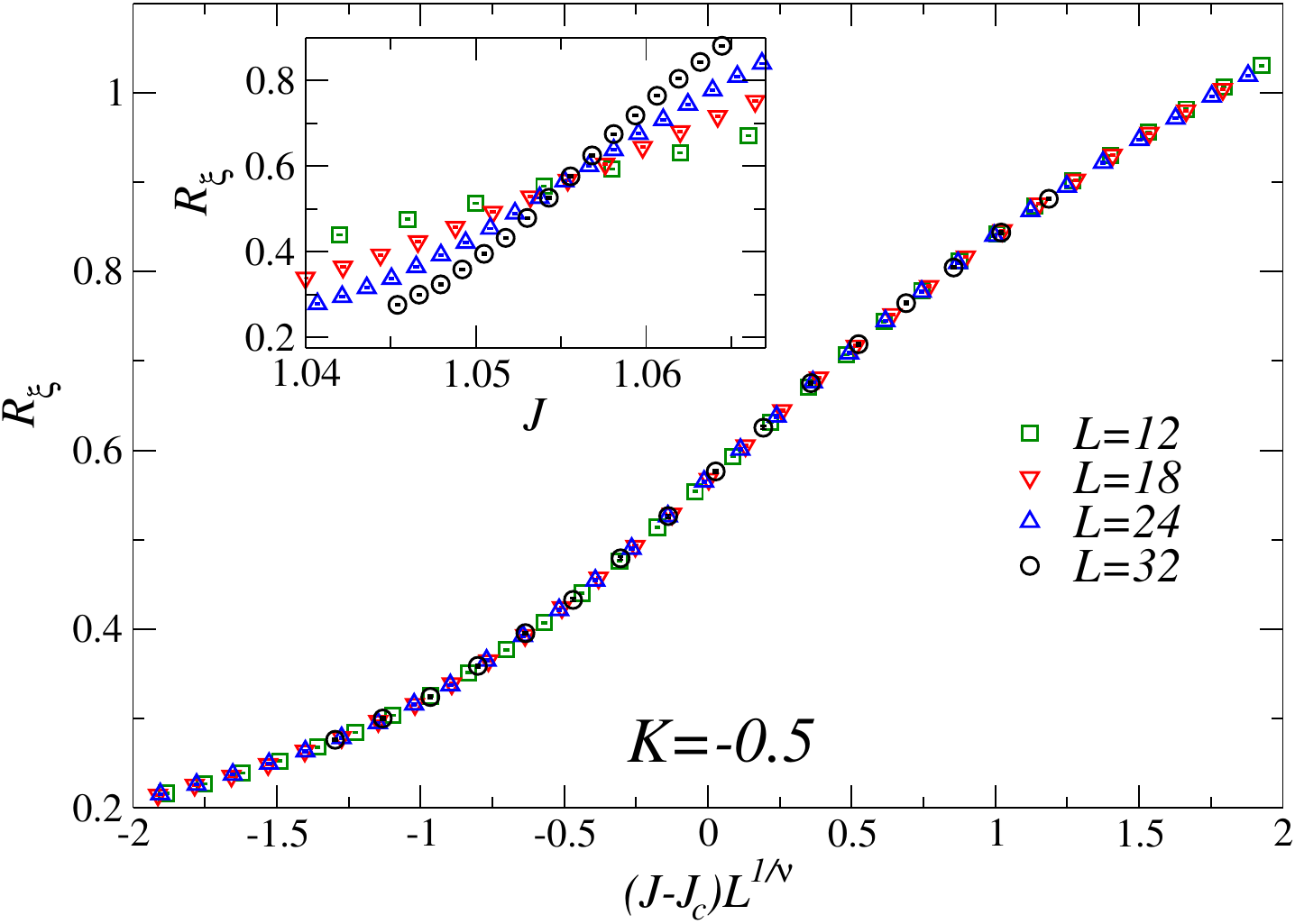}
  \caption{Some results of the MC simulations of the single-charge
    $N=2$ CLAH model along the line $K=-0.5$ for several values of
    $J$, up to $L=32$.  The bottom panel shows data of $R_{\xi}$
    versus $(J-J_c)L^{1/\nu}$, with $\nu=0.71164$, which is the value
    of the length-scale critical exponent of the 3D O(3) universality
    class~\cite{Hasenbusch-20,
      Chester-etal-20-o3,KP-17,HV-11,CHPRV-02,GZ-98}, and our estimate
    of the transition point, $J_c=1.05535$, from the crossing point of
    the data versus $J$, see the inset.  The top panel shows a plot of
    the data of the Binder parameter $U$ versus $R_{\xi}$, comparing
    them with the asymptotic universal curve for the O(3) universality
    class (solid line), obtained in Ref.~\cite{BPV-20-hcAH} with a
    relative accuracy of at most about $0.5\%$ (this uncertainty
    corresponds to the solid lines above and below the central one).
  \label{km0p5data}}
\end{figure}

The FSS analyses at both values of $K$ provide clear evidence of
continuous transitions belonging to the 3D O(3) universality class,
like the transition line for $K\ge 0$. Indeed, the universal critical
behavior, i.e., critical exponents and FSS functions, are in full
agreement with the most accurate results obtained for the 3D O(3)
vector universality class, see, e.g.,
Refs.~\cite{PV-02,Hasenbusch-20,BPV-20-hcAH,Chester-etal-20-o3,
  KP-17,HV-11,CHPRV-02,GZ-98}. In particular, the data of the ratio
$R_\xi$, see Eq.~\eqref{rxi}, show a clear crossing point, see for
example bottom Fig.~\ref{km0p5data} for the case $K=-0.5$ (analogous
results are obtained for $K=-0.7$, not shown), with slopes that are
fully consistent with the most accurate estimate~\cite{Hasenbusch-20}
$\nu= 0.71164(10)$ of the length-scale critical exponent of the 3D
O(3) vector universality class.  Then, quite accurate estimates of the
transition points are obtained by matching these data with the FSS
behavior (see, e.g., Refs.~\cite{ZJ-book,PV-02,BPV-25})
\begin{equation}
  R_\xi = f_\xi(X) + O(L^{-\omega}),\qquad X = (J-J_c)L^{1/\nu},
  \label{rxifss}
\end{equation}
using the O(3) value for $\nu$, and the O(3) scaling-correction
exponent~\cite{Hasenbusch-20} $\omega=0.759(2)$.  By straightforward
fits using polynomial approximations of $f_\xi(X)$ around the crossing
point $J_c$, we obtain $J_c= 1.05535(15)$ for $K=-0.5$ and
$J_c=1.1760(3)$ for $K=-0.7$. We also mention that an analogous
analysis of the Binder parameter $U$ gives $J_c= 1.0555(3)$ and
$J_c=1.1756(4)$ for $K=-0.5$ and $K=-0.7$, respectively, in full
agreement with the results obtained from $R_\xi$.  Using these
estimates of $J_c$ and the O(3) value of $\nu$, the data of $R_\xi$
versus $X$ show a good scaling, consistent with
Eq.~(\ref{rxifss}), see the bottom of Fig.~\ref{km0p5data} for the
case $K=-0.5$. The scaling of the data for $K=-0.7$ is completely
analogous and not shown. Moreover, as reported in the top panel of
Fig.~\ref{km0p5data}, the data for the Binder parameter $U$ versus the
ratio $R_\xi$ show the expected universal FSS behavior (see, e.g.,
Refs.~\cite{BPV-20-hcAH,BPV-25,BMV-26})
\begin{equation} 
  U = F_U(R_\xi) + O(L^{-\omega})
  \label{uvsrxi}
\end{equation}
in good agreement with the scaling function $F_U(R_\xi)$ computed for
the standard O(3) vector model in Ref.~\cite{BPV-20-hcAH}.  Also data
for $K=-0.7$ (not shown) approach the asymptotic O(3) scaling function
$F_U(R_\xi)$, with only slightly larger scaling corrections than for
$K=-0.5$.

In conclusion these FSS analyses up to $L=32$ show clearly that the
line of O(3) continuous transitions crossing the $K=0$ axis continues
at least down to $K=-0.7$, where the average plaquette is already
negative, indeed $W \approx -0.39$ (for $K=-0.5$ we find $W\approx
-0.21$), thus persisting also in the presence of a moderately weak
frustration.

\subsection{The first-order transition line}
\label{firstorder}

We now present numerical finite-size analyses at some fixed values of
$K\lesssim -0.7$, in particular at $K=-0.8$, $K=-1$ and $K=-1.5$,
which provide evidence of a first-order transition line for $K\lesssim
-0.7$, which blends into the O(3) transition line discussed in
Sec.~\ref{conttra}, and which extends down to $K=-\infty$.  Like the
continuous transitions reported in Sec.~\ref{conttra}, this
first-order transition line $J_{\rm fo}(K)$ separates the disordered
phase (low $J$) from the ordered phase (high $J$) where the global
SU(2) symmetry is spontaneously broken, due to the condensation of the
bilinear operator $Q_{\bm x}$, cf. Eq.~(\ref{qdef}). Indeed, for
$J>J_{\rm fo}(K)$ we observe the asymptotic finite-size behaviors
expected within a phase characterized by the spontaneous breaking of a
continuous symmetry, such as~\cite{GL-87,HL-90}
\begin{equation}
  \chi_Q = c\,V \left[1 + O(L^{-1})\right],\quad
  U = 1 + O(L^{-2}).
    \label{asybropha}
\end{equation}

\begin{figure}[tbp]
 \includegraphics*[width=0.90\columnwidth]{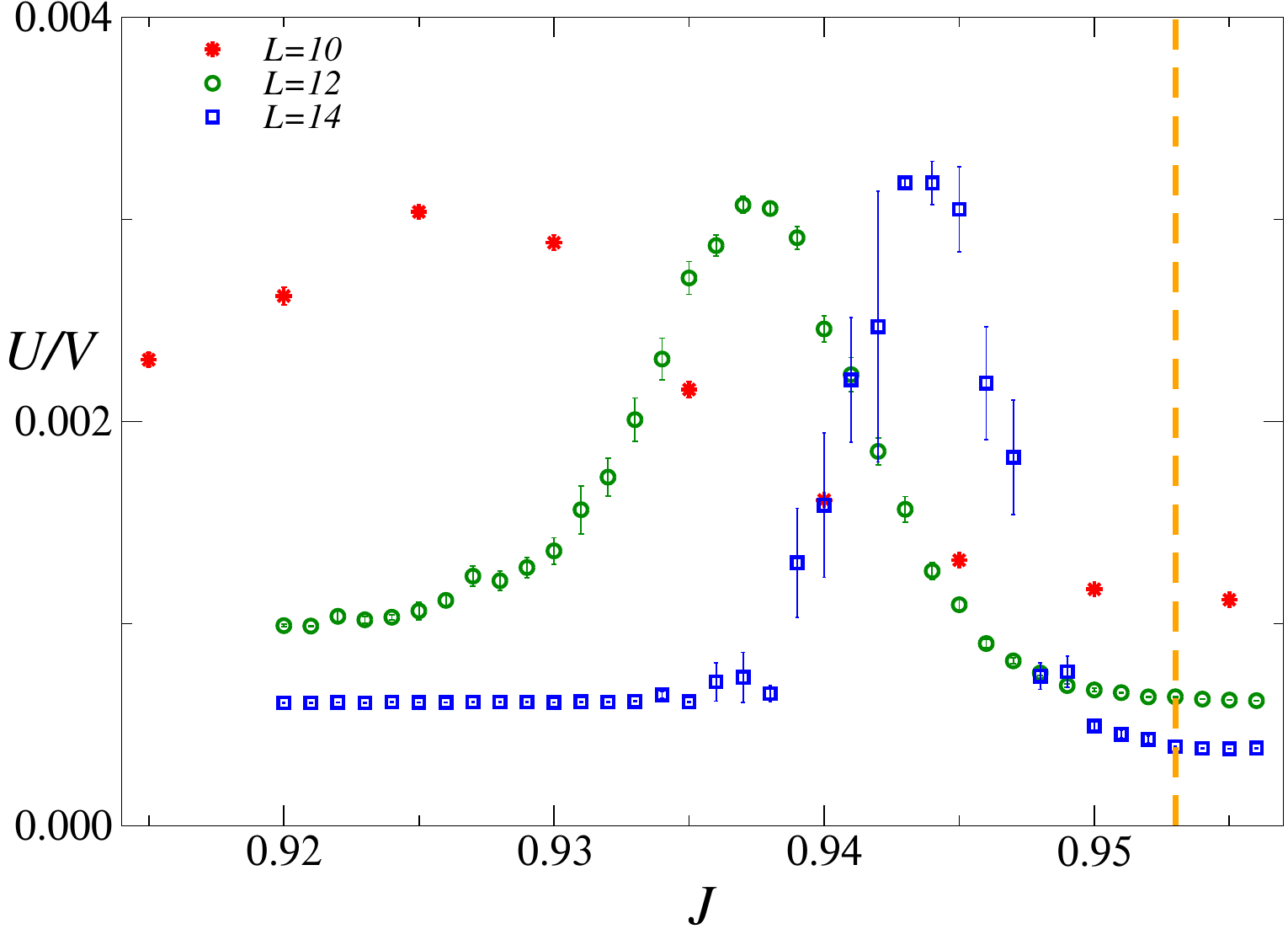}
 \includegraphics*[width=0.90\columnwidth]{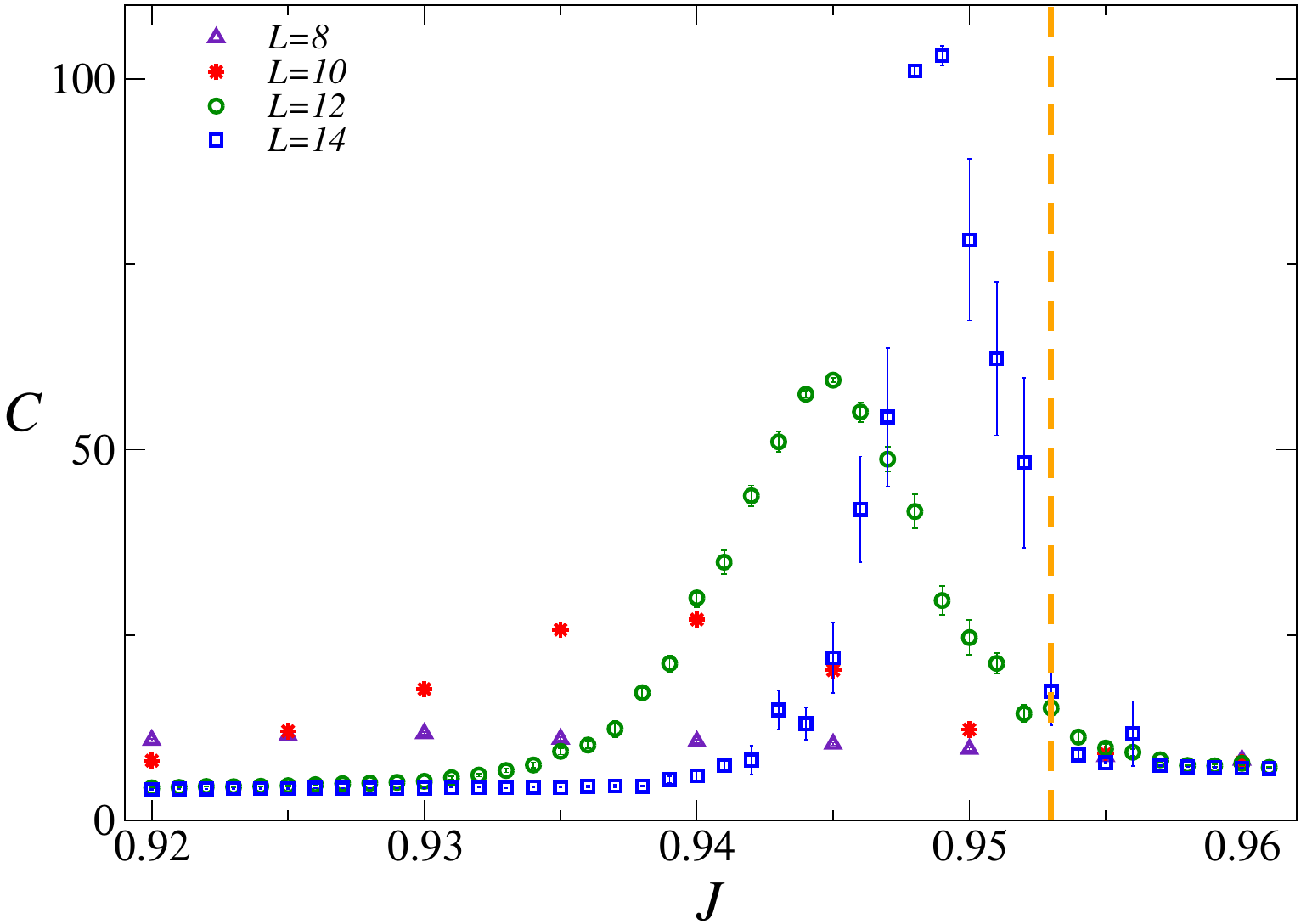}
 \caption{The specific heat $C$ (bottom) and the Binder parameter $U$
   (top) versus $J$, for $K=-1$ and lattice sizes up to $L=14$ in the
   single-charge $N=2$ model.  In the case of $U$, we plot the ratio
   $U/V$ to highlight the large-size behavior $U_{\rm max}\sim V$ of
   its peak, cf. Eq.~(\ref{cvbifo}).  The vertical dashed line
   indicates our best estimate of the transition point $J_{\rm
     fo}=0.953(2)$.}
   \label{km1data}
\end{figure}

Numerical evidence for these first-order transitions can be obtained
by examining the energy-density distribution, the specific heat $C$,
and the Binder parameter $U$ as functions of $J$ at fixed $K$. At the
first-order transition point, $J_{\rm fo}$, the energy-density
distribution is expected to develop two well-separated peaks, whose
separation provides an estimate of the latent heat
$\Delta_h$. Moreover, in finite-size systems, both the specific heat
and the Binder parameter exhibit peaks as $J$ is varied, with their
peak heights diverging proportionally to the volume $V=L^3$ close to
the transition point $J_{\rm fo}$ separating an ordered from a
disordered phase.  More precisely, their peak values behave
as~\cite{Binder-87,PV-24}
\begin{equation}
  C_{\rm max}(L) = {\Delta_h^2\over 4} V+O(1),\quad
  U_{\rm max}(L) = a\,V + O(1),
    \label{cvbifo}
\end{equation}
and their peaks are located at
\begin{equation}
  J_{\rm max}^{C/U}(L) \approx J_{\rm fo} + \frac{b_{C/U}}{V},
  \label{jpecu}
\end{equation}
which allows us to estimate $J_{\rm fo}$ (note that the coefficients
$b_{C/U}$ of the subleading terms are not the same for $C$ and $U$).

For example, in Fig.~\ref{km1data} we show data for the specific heat
$C$ and Binder parameter $U$ up to $L=14$ for $K=-1$. They show clear
peaks whose maxima rapidly increase with the lattice size, roughly
proportionally to the volume, as predicted by Eq.~(\ref{cvbifo}).
From the size dependence of the well separated peaks in the energy
distribution (not shown) and the maxima of the specific heat, see the
bottom Fig.~\ref{km1data}, we estimate $\Delta_h = 0.42(1)$.  The
large-$L$ extrapolations of the positions of the peaks of the specific
heat and Binder parameter, obtained by fitting them using
Eq.~(\ref{jpecu}), give $J_{\rm fo} = 0.9537(13)$ and $b=-14(3)$ from
$C$, and $J_{\rm fo} = 0.9529(16)$ and $b=-26(3)$ from $U$, with
acceptable $\chi^2/{\rm d.o.f}\lesssim 1$. Thus, we consider $J_{\rm
  fo} = 0.953(2)$ as our best estimate of the transition point at
$K=-1$.

Analogous results have been obtained at the other values of $K$ for
which we have performed MC simulations, i.e., $K=-0.8$ and $K=-1.5$,
leading to the estimates $J_{\rm fo} = 1.115(8)$ and $\Delta_h \approx
1.0$ for $K=-0.8$ (in this case the first-order transition is so
strong that we can reliably obtain results up to sizes $L=8$ only),
and $J_{\rm fo} = 0.860(5)$ and $\Delta_h \approx 0.08$ for $K=-1.5$.

We mention that also the average plaquette $W$ appears to be
discontinuous at these first-order transitions. In particular, at
$K=-0.8$ the average plaquette passes from $W\approx -0.56$ to
$W\approx -0.64$ when increasing $J$ across $J_{\rm fo}\approx 1.115$
(note that $W$ is nonmonotonous, because it decreases across the
transition, and then it starts growing again with increasing $J$), and
at $K=-1$ it passes from $W \approx -0.75$ to $W\approx -0.77$ across
$J_{\rm fo}\approx 0.95$.

More accurate estimates of the transition points $J_{\rm fo}$ and of
latent heat at these transitions would require the use of MC
algorithms specifically designed to study strong first-order
transitions, such as those discussed in
Refs.~\cite{BN-92,WL-01,LLR-12,Janke-03}. But this is beyond the scope
of our numerical investigation.

Summarizing, these results show that the continuous O(3) transition
line for $K\gtrsim -0.7$ turns into a first-order transition line
below a finite negative value $K^*$, i.e., for $K<K^*$ with
$K^*\lesssim -0.7$, with a nonmonotonic latent heat that rapidly
increases up to a maximum for $K\approx -0.8$, and then decreases
toward large negative values of $K$. Note that the average plaquette
$W$ is always negative along the first-order transition line, and it
displays a nonmonotonic behavior in $J$ at fixed $K$.

\subsection{The fully-frustrated $K\to-\infty$ limit}
\label{fflimit}

The rapid drop of the latent heat when decreasing $K$ within the
region $K\lesssim -1$ leads to the natural question whether the
first-order transition line persists down to $K\to -\infty$, or turns
into a continuous transition at a finite value of $K$. Some additional
information can be obtained by analyzing the CLAH in the limit $K\to
-\infty$, where the average plaquette approaches the value $W\to -1$,
leading to a fully frustrated system of complex spin variables ${\bm
  z}_{\bm x}$ (apart from gauge transformations), for example by
fixing the gauge configuration (\ref{fruconf}).  The resulting model
is equivalent to a fully-frustrated simple-cubic O(4) vector model,
whose Hamiltonian is reported in Eq.~(\ref{homfru}).

\begin{figure}[tbp]
 \includegraphics*[width=0.90\columnwidth]{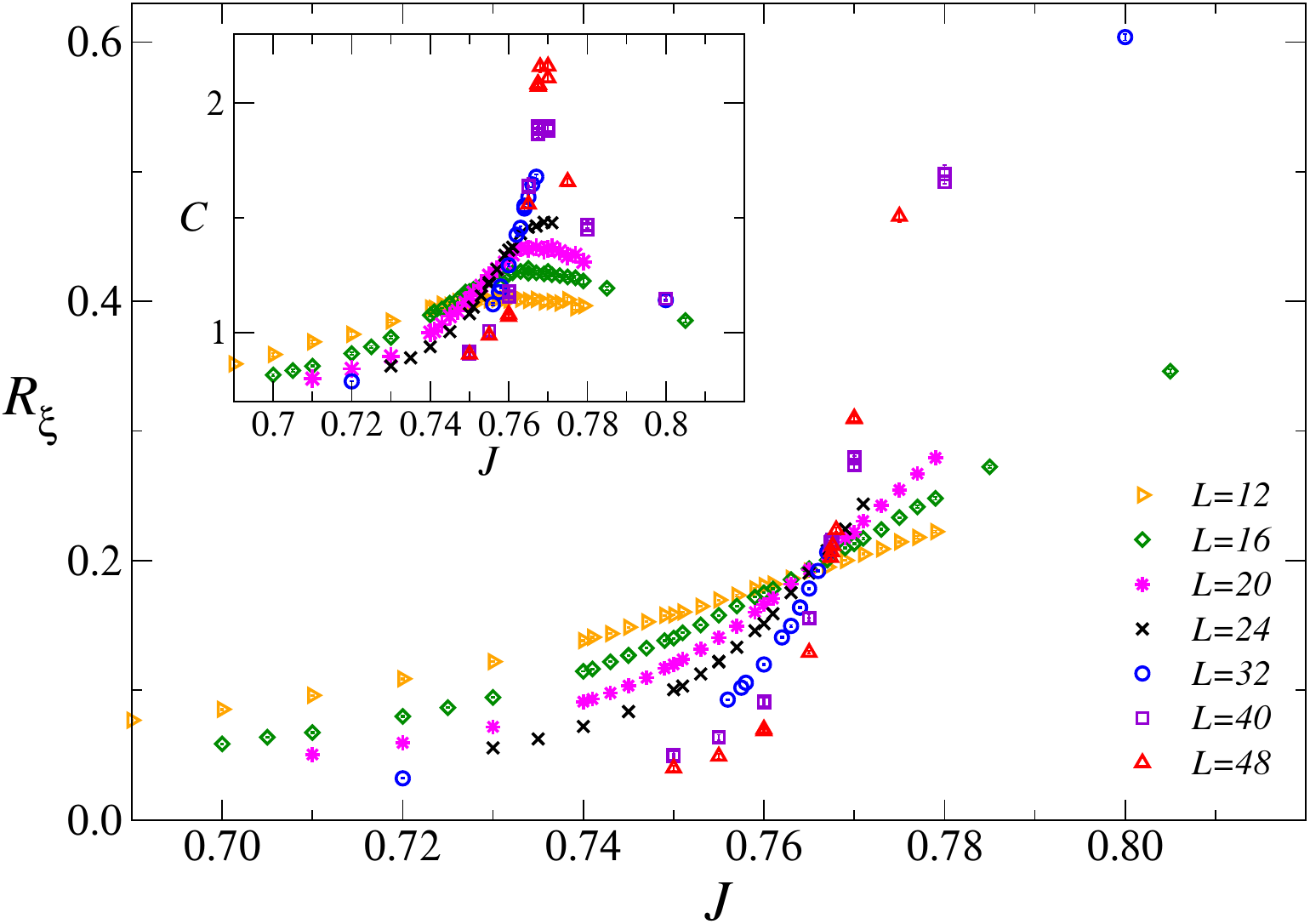}
 \caption{The ratio $R_\xi$ and specific heat (in the inset) for the
   fully frustrated spin model obtained by the $K\to-\infty$ limit of
   the single-charge $N=2$ CLAH model, for various lattice sizes up to
   $L=48$.  The data show a peak in the specific heat and a rapid
   variation of $R_\xi$ around $J=0.77$, with a quite stable crossing
   point of the data of $R_\xi$ for different sizes, indicating the
   occurrence of a phase transition.
   \label{ffrudata}}
\end{figure}

To investigate the nature of the transition for the fully-frustrated
simple-cubic O(4) model, we present a FSS analysis of the $N=2$ CLAH
model fixing the link variables to their $K\to -\infty$ limit
(\ref{fruconf}), corresponding to a particular gauge fixing.  Our
numerical simulations up to $L=48$ show evidence of a phase transition
at $J=J_{\rm tr}\approx 0.77$, as can be seen from
Fig.~\ref{ffrudata}, showing data for the ratio $R_\xi\equiv \xi/L$,
where $\xi$ is second-moment correlation length associated with
the two-point function of the bilinear operator $Q_{\bm x}^{ab}$, and
the specific heat in its inset.  Indeed, at $J_{\rm tr}\approx 0.77$
the specific heat shows a peak whose maximum appears to increase with
increasing $L$, and the finite-size behavior of $R_\xi$ shows a rapid
variation of the data for various lattice sizes, with a relatively
stable crossing point.  Both features are characteristic of the
finite-size behavior at a phase transition.  However, a more
quantitative analysis is required to assess the nature of the
transition, whether it is continuous or first order.  In particular,
at first-order transitions, peculiar finite-size effects arise from
the coexistence of the two phases~\cite{PV-24} (when the boundary
conditions do not favor either phase, such as periodic boundary
conditions).

Analogously to what has been observed for a smaller number of
components~\cite{NHD-10,NHD-11}, i.e., $M=2$ and $M=3$, the data at
small lattice sizes apparently favor a continuous transition. However,
signs of a weak first-order transition are eventually observed by
looking at the behavior for larger lattices. Indeed, although the
lattice sizes considered, up to $L=48$, are not sufficiently large to
reveal a double-peak structure in the energy-density distribution or
the corresponding volume scaling of the specific heat, the behavior of
the Binder parameter $U$ suggests that the transition is ultimately of
first order.

\begin{figure}[tbp]
 \includegraphics*[width=0.90\columnwidth]{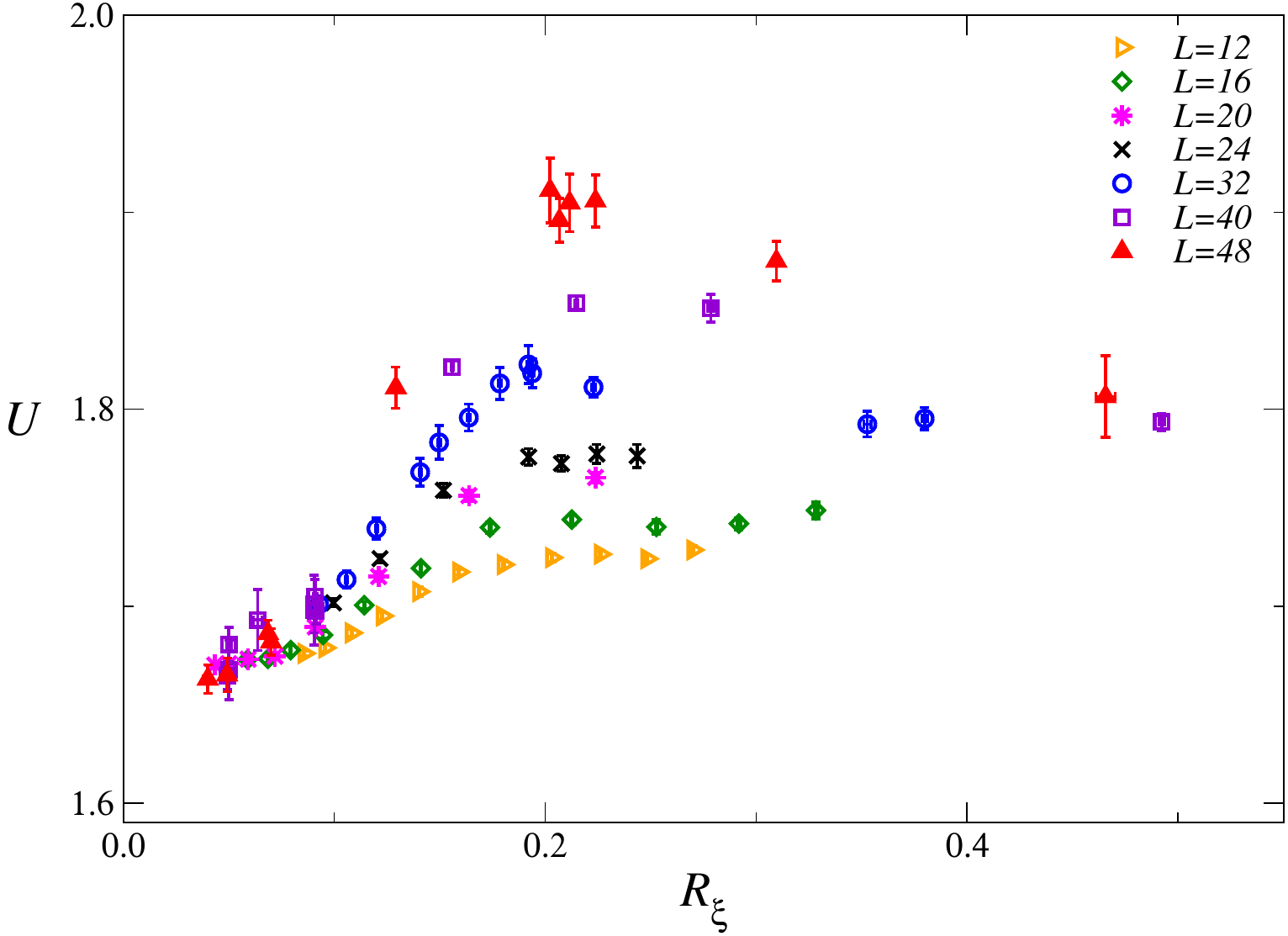} 
 \caption{The Binder parameter $U$ versus $R_\xi$ for the
   fully-frustrated spin model obtained by taking the $K\to-\infty$
   limit of the single-charge $N=2$ CLAH model.
  \label{ffrudataU}}
\end{figure}

We recall that, in the case of a continuous transition driven by the
condensation of $Q_{\bm x}^{ab}$, the data of the corresponding Binder
parameter $U$ versus $R_\xi$ are expected to converge to a universal
function in the large-$L$ limit, as in Eq.~(\ref{uvsrxi}).  However,
as shown in Fig.~\ref{ffrudataU}, no scaling behavior is observed in
the present case.  Instead, the data reveal the emergence of a peak
whose height appears to increase with increasing $L$.  This is usually
considered as an early indicator of a weak first-order transition,
suggesting the divergence of the Binder parameter in the limit
$L\to\infty$~\cite{PV-24,BMV-26,CPPV-04}.\footnote{It is also worth
noting that the Binder parameter goes from the O(4) value $U=5/3$
within the disordered phase (5/3 is the value of the Binder parameter
associated to the tensor operator $Q_{\bm x}$ in the disordered phase
of the vector O(4) model~\cite{PV-19-AH3d}) to a nontrivial value,
$U\approx 1.79$, which is far from the value $U=1$ of standard ordered
ferromagnetic phases.  Note also that the value $U=1$ is eventually
expected in the CLAH model for any finite $K<0$, because the model
always orders in the limit $J\to\infty$. This is not necessarily true
for the fully-frustrated O($M$) models, because they is formally
obtained by first taking the limit $K\to -\infty$ of the CLAH model.
We find that $U$ converges to a nontrivial value for $J> J_{\rm
  tr}\approx 0.77$.  For example, we have obtained $U=1.78(2)$ at
$J=1.4$ for $L=24$, $U=1.787(10),\,1.796(8)$ at $J=0.8,\,1.0$ for
$L=32$, and $U=1.78(4),\,1.79(2)$ at $J=0.8,\,1.0$ for $L=40$.  The
nontrivial value $U\approx 1.79$ suggests that full frustration gives
rise to complex energy-minimizing configurations.  Actually, this also
suggests that the properties of the low-temperature phase do not
converge uniformly in the limit $K\to -\infty$, because we recall that
in the limit $J\to\infty$ we must have $U\to 1$ for any $K$.}

Further evidence in favor of a first-order transition comes from the
behavior of the effective critical exponent obtained by fitting the
data for the ratio $R_\xi$ to the FSS form (\ref{rxifss}). These fits
yield a rather stable estimate of the crossing point, $J_c=0.768(1)$,
but the effective exponent $\nu_{\rm eff}$ decreases significantly as
the minimum lattice size $L_{\rm min}$ included in the fits
increases. In particular, fits restricted to the largest lattice
sizes, corresponding to $L_{\rm min}=32,40$, typically yield
$\nu_{\rm eff}\lesssim 0.5$, which is incompatible with a continuous
transition~\cite{PV-26}.

In conclusion, we believe that the above results favor a first-order
transition in the $K\to -\infty$ limit.  Therefore, a natural
hypothesis is that the transition line starting from $K^*\lesssim
-0.7$ remains first order for all $K<K^*$, up to $K=-\infty$.

Concerning the fully frustrated O($M$) vector models, we conclude that
the critical number of components $M^*$, at which the transition
changes from first order to continuous, must be larger than 4. We
believe that this issue deserves further investigation, in particular
to determine whether $M^*$ is finite, i.e., whether the continuous
transition found in the fully-frustrated spherical
limit~\cite{DPTV-80} persists at finite values of $M$.

\subsection{Topological transitions within the high-$J$ phase}
\label{toptra}

\begin{figure}[tbp]
  \includegraphics*[width=0.90\columnwidth]{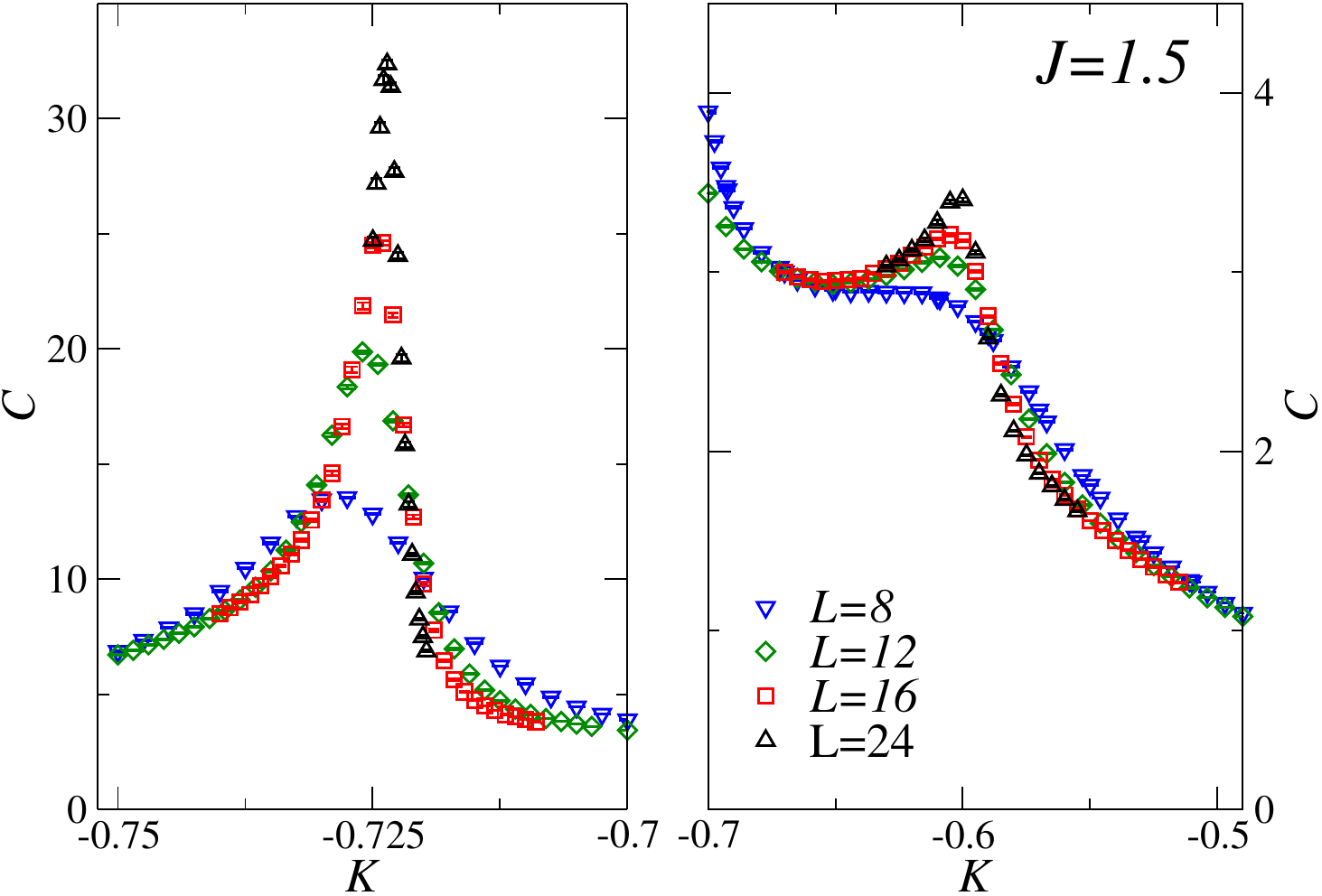}
  \caption{Specific heat of the single-charge $N=2$ model at $J=1.5$
    as a function of $K$.  Two different transitions can be noted,
    quite close to each other. Note the different scales in the two
    panels.
  \label{N2_C}}
\end{figure}

\begin{figure}[tbp]
  \includegraphics*[width=0.90\columnwidth]{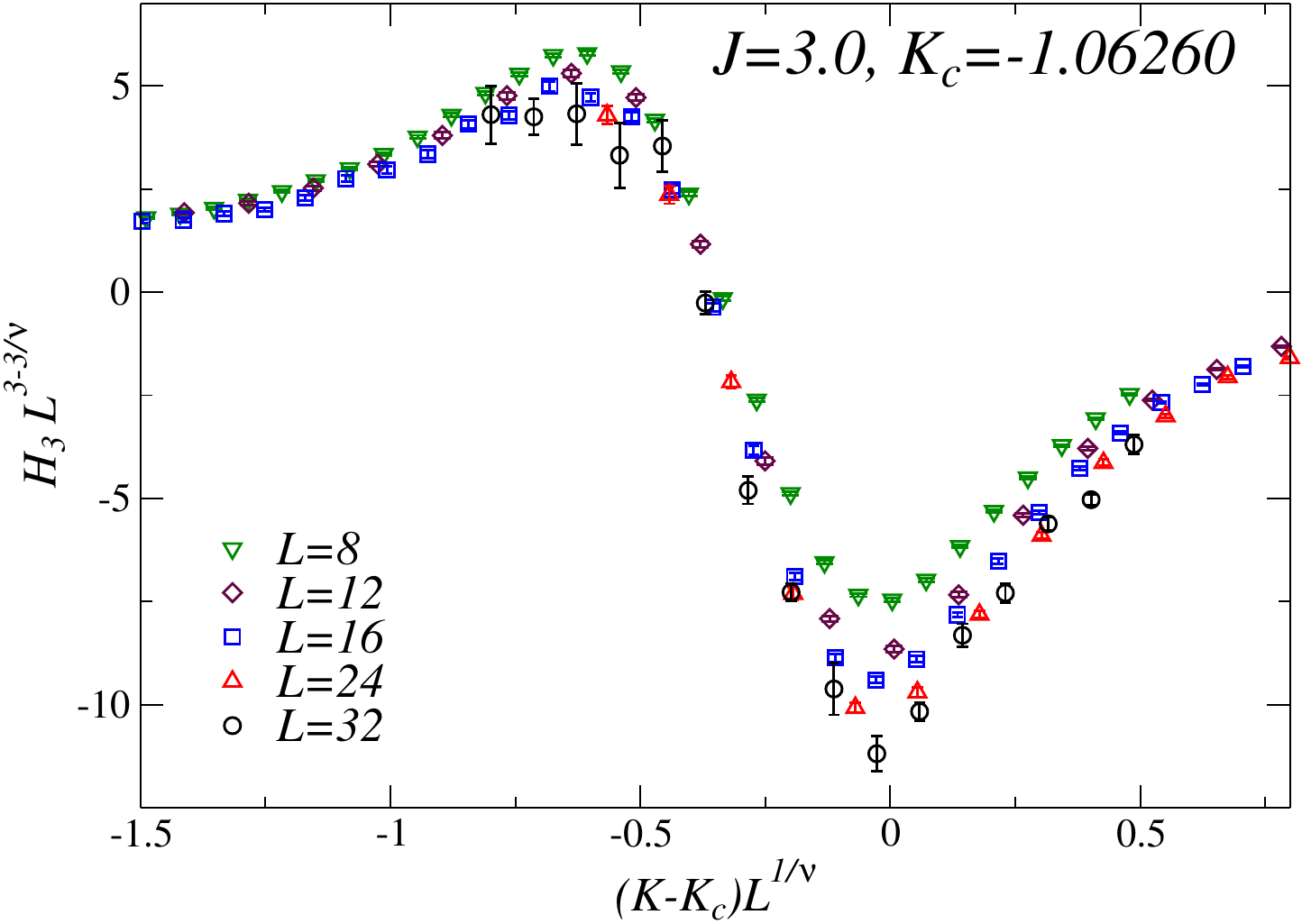}
  \includegraphics*[width=0.90\columnwidth]{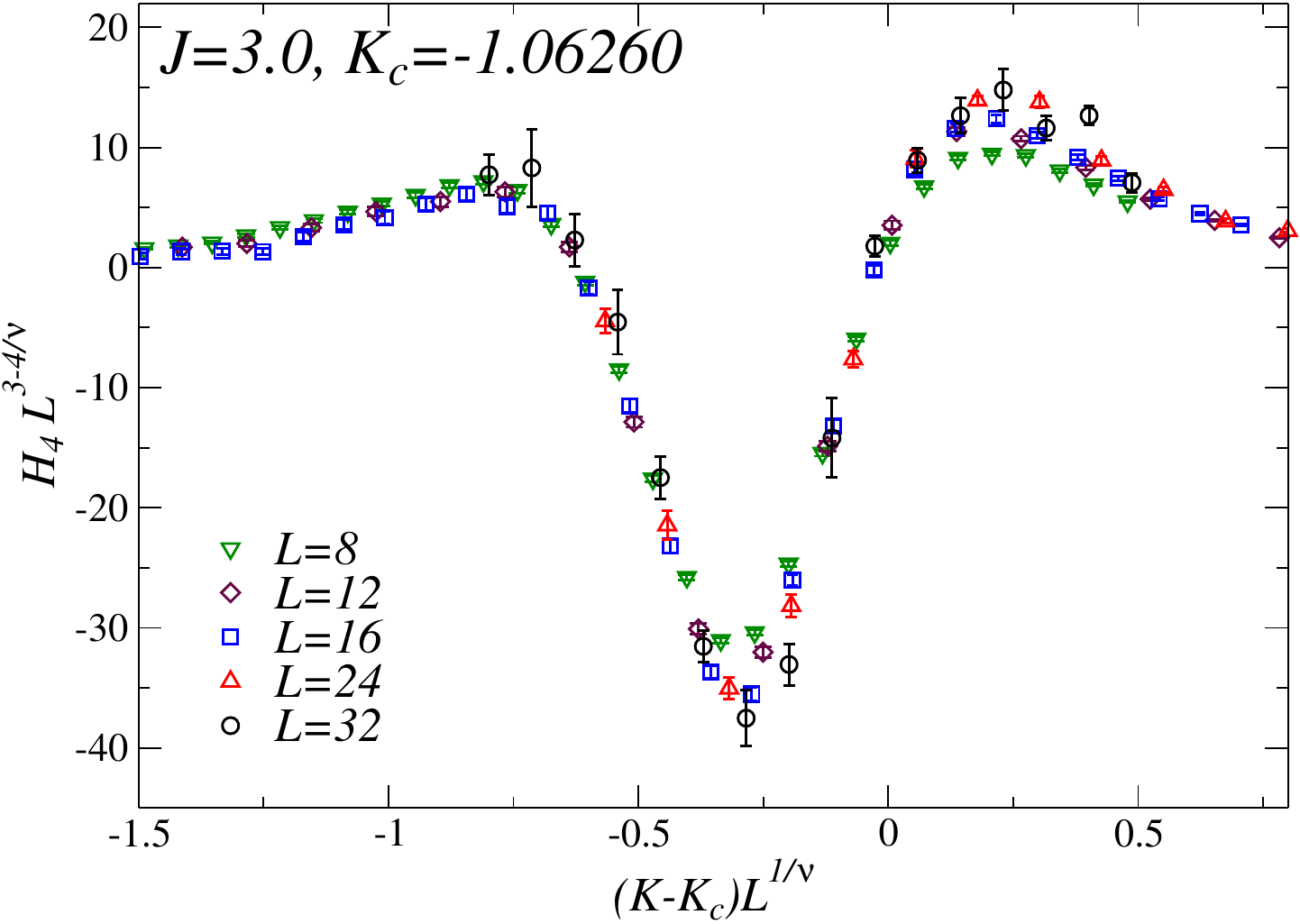}
  \caption{Scaling of the third and fourth cumulants of the energy for
  the single-charge $N=2$ model at $J=3$ for $K\approx -1.06$, obtained
  by fixing the critical exponent $\nu$ to the value of the 
  three-dimensional Ising model.
  \label{N2_ising}}
\end{figure}

\begin{figure}[tbp]
  \includegraphics*[width=0.90\columnwidth]{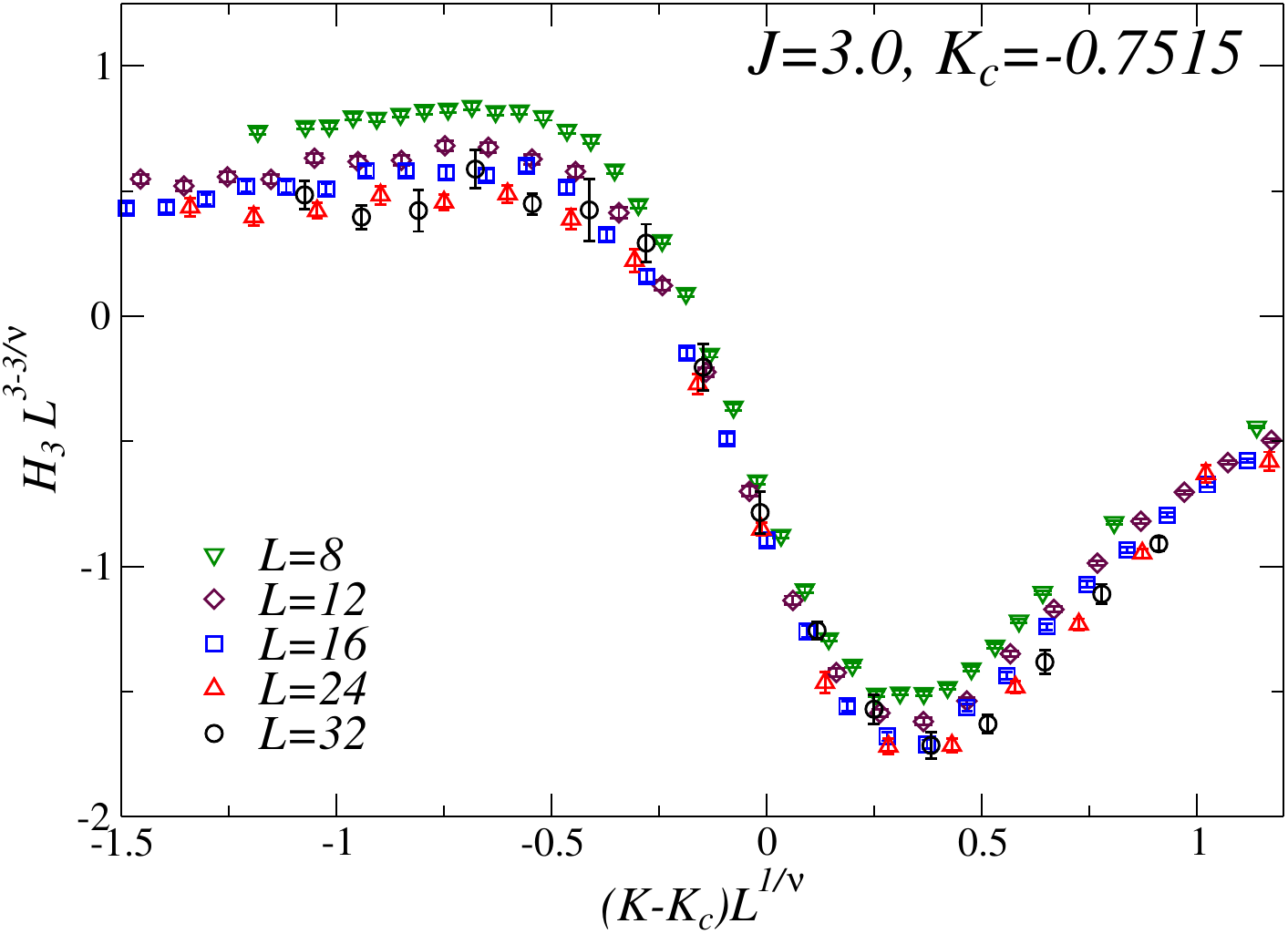}
  \includegraphics*[width=0.90\columnwidth]{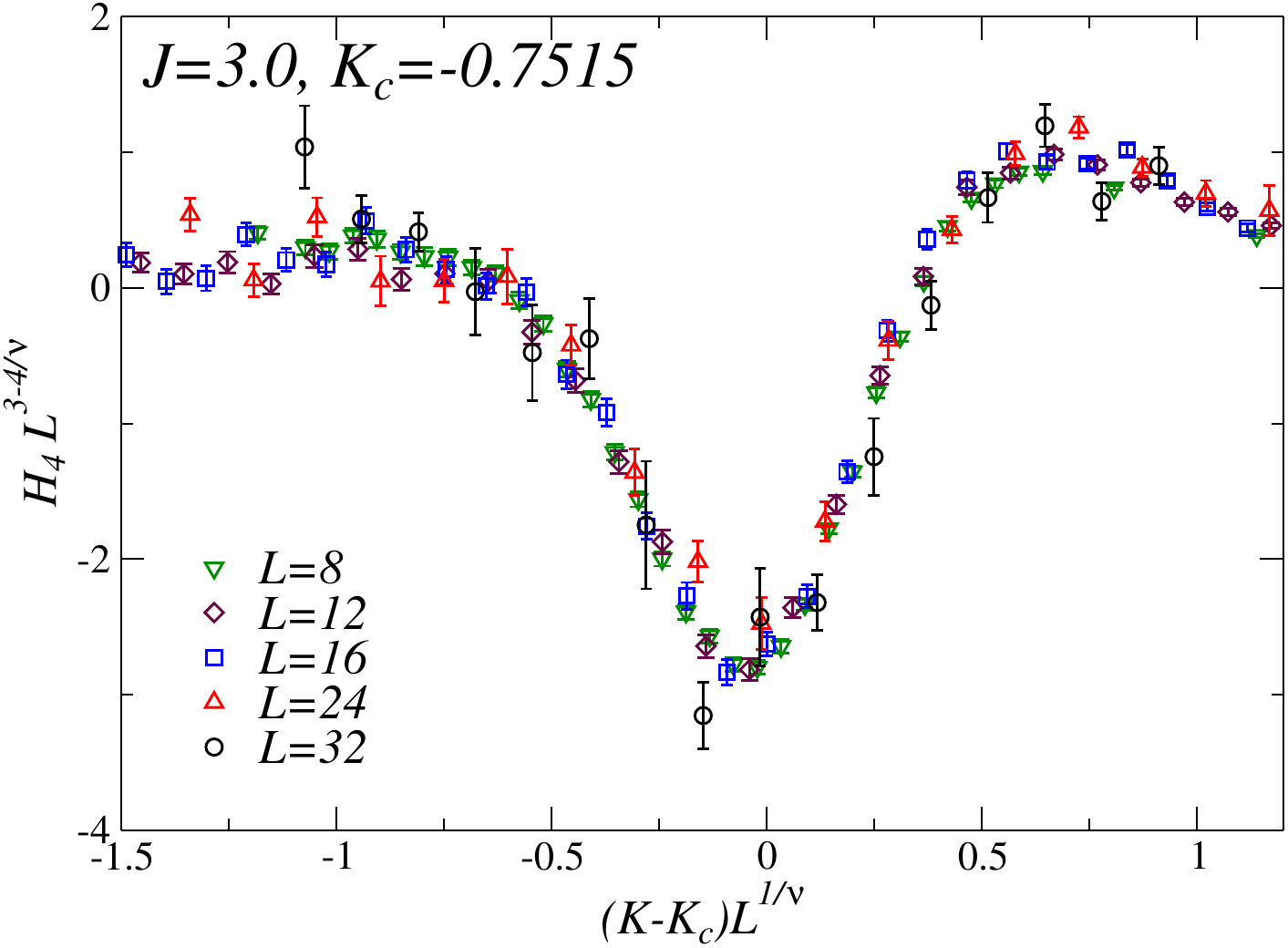}
  \caption{Scaling of the third and fourth cumulants of the energy for
  the single-charge $N=2$ model at $J=3$ for $K\approx -0.75$, obtained
  by fixing the critical exponent $\nu$ to the value of the 
  three-dimensional XY model.
  \label{N2_xy}}
\end{figure}

When discussing the behavior of the plaquette $W$ shown in
Fig.~\ref{Wfigs}, we noted hints of additional transition lines beyond
those expected on the basis of the global SU(2) symmetry of the $N=2$
CLAH model. To clarify this point, we present numerical results at
$J=1.5$, just above the point where the O(3) and discontinuous
transition lines meet, and at $J=3.0$, significantly further away from
this point in the phase diagram. In both cases the susceptibility
$\chi_Q$ and the Binder cumulant $U$ of the SU(2) order parameter turn
out to scale as expected in a phase characterized by the spontaneous
breaking of a continuous symmetry, i.e., according to
Eq.~\eqref{asybropha}. Therefore, in both cases we are well within the
SU(2) broken phase.

Results for the specific heat at $J=1.5$ are reported in
Fig.~\ref{N2_C}. The left panel shows a pronounced peak at $K\approx
-0.725$, while the right panel reveals another (much less pronounced)
peak at $K\approx -0.6$ (note the different scales in the two panels),
superimposed on the right tail of the former. Analogous results are
found for $J=3$, with peaks at $K\approx -1.06$ and $K\approx -0.75$.
These data strengthen our hypothesis of the existence of two
transition lines within the broken high-$J$ phase.

Apparently, the transitions at both $J=1.5$ and $J=3$ do not have any
local order parameter, suggesting that they have a topological nature
related to the correlations of the U(1) gauge variables.  Therefore,
to investigate their nature, we resort to the FSS behavior of the
energy-density cumulants defined in Eqs.~\eqref{H3} and
\eqref{H4}. Close to a continuous phase transition these cumulants
develop a critical behavior governed by the asymptotic FSS
relations~\cite{SSNHS-03, BPV-24-coH,BPV-25}
\begin{equation}
H_k \approx L^{k/\nu-3} {\cal H}_k(X) \left[1 + L^{-\omega}
    {\cal H}_{k,\omega}(X)\right] + b_k, \label{H3-scaling}
\end{equation}
where $X = (K-K_c)L^{1/\nu}$, the constants $b_k$ represent the
analytic background, and the $O(L^{-\omega})$ suppressed term (with
$\omega>0$) represents the leading scaling correction. The functions
${\cal H}_k$ and ${\cal H}_{k,\omega}$ are universal (up to an overall
rescaling and a rescaling of their arguments), depending only on the
universality class of the transition and on other infrared features,
such as the boundary conditions adopted in the simulations. The
contributions from analytic background terms become progressively less
important as the cumulant order $k$ increases. However, the
statistical errors of $H_k$ also increase with $k$. We therefore
restrict our analysis to the third and fourth cumulants.

Since for $J=1.5$ the two transitions are very close to each other,
determining their universality classes is particularly challenging, as
significant crossover effects are expected. For this reason, we report
a FSS analysis only for the data at $J=3$, where the results appear
more transparent.

To extract the values of $\nu$ and $K_c$ using Eq.~\eqref{H3-scaling},
we performed several fits, approximating the unknown scaling functions
${\cal H}_k$ and ${\cal H}_{k,\omega}$ by polynomials. Although
scaling corrections are visible in our data, the statistical accuracy
is insufficient to determine both the correlation length critical
exponent $\nu$ and the leading correction to scaling exponent
$\omega$.  We therefore treated $\nu$ as a free parameter and repeated
the fits for different fixed values of $\omega$ in the range
$0.5\leq\omega\leq 1$. Applying this analysis to the third cumulant
$H_3$ at $J=3.0$, we obtain $K_c=-1.06260(15)$ and $\nu=0.625(15)$ for
the first transition, and $K_c=-0.7510(5)$ and $\nu=0.685(15)$ for the
second one. The quoted uncertainties include systematic effects
associated with the choice of $\omega$, the polynomial approximation
of the scaling functions, the fitting range, and the minimum lattice
size $L$ included in the fits. An analogous analysis of the fourth
cumulant $H_4$ yields $K_c=-1.06260(5)$ and $\nu=0.62(1)$ for the
first transition, and $K_c=-0.7514(4)$ and $\nu=0.675(10)$ for the
second one.

In the absence of a clear theoretical framework for interpreting these
two continuous transition lines, it is natural to conjecture that the
first transition (occurring at smaller, more negative values of $K$)
belongs to the ${\mathbb Z}_2$ gauge universality class, while the
second belongs to the XY gauge universality class. They share the same
critical exponents as the standard Ising and XY universality classes,
although they may be considered as distinct universality classes with
their own peculiarities, in particular, they are characterized by the
absence of local order parameters, see, e.g., Ref.~\cite{BPV-25} for a
discussion of this point. Indeed, for the Ising universality class,
the most accurate estimate is $\nu_I=0.62997097(12)$
from~\cite{Chang-etal-25} (and $\omega_I=0.8295(6)$
from~\cite{Reehorst-22}); see also~\cite{PV-02, GZ-98, CPRV-02, KP-17,
  DBTW-20, Hasenbusch-21} for other determinations. For the XY
universality class, we have instead $\nu_{\rm XY}=0.671718(23)$
from~\cite{Hasenbusch-25} (and $\omega_{\rm XY}=0.789(4)$
from~\cite{Hasenbusch-19}), see also~\cite{GZ-98, KP-17, DBTW-20,
  Chester-et-al-20, CHPV-06,BMPS-06} for related results.  In
Figs.~\ref{N2_ising} and \ref{N2_xy}, we show the quality of the data
collapse obtained by fixing the critical exponent $\nu$ to its most
accurate available estimate for the Ising and XY universality classes,
respectively.

A comparison of the data for the fourth energy cumulant reported in
Fig.~\ref{N2_ising} with the corresponding ones obtained by
simulating the 3D $\mathbb{Z}_2$ gauge model with periodic boundary
conditions (analogous to that carried out in \cite{BV-26-z2h}) shows
substantial agreement, but a proper extrapolation to $L\to\infty$ of
both datasets would be required to draw definite conclusions from this
comparison.
 
Since no examples are known of 3D systems whose critical exponents
vary continuously along a transition line, it is finally natural to
expect the same critical exponents of the Ising and XY universality
classes to characterize the entire topological transition lines.  The
results for $J=1.5$ turn out to be substantially consistent with this
scenario, although they appear less stable, likely due to crossover
effects that are not solved up to the available maximum lattice size.

\section{Numerical results for the one-component CLAH model}
\label{numresN1}

As already mentioned, the phase diagram of the $N=1$, $q=1$ model is
trivial for $K\ge 0$, consisting of a single phase with no transition
lines. We now present a numerical investigation of the model at
negative gauge coupling, aimed at understanding how the phase diagram
extends to $K<0$ and whether additional phases emerge.

We know that, in the limit $K\to -\infty$, where the model becomes
equivalent to the fully-frustrated XY model, a first-order transition
occurs at a finite value of $J$, namely $J_{\rm fo}\approx
0.736$~\cite{NHD-10}. Since first-order transitions are generally
robust against perturbations, this transition is expected to persist
at finite values of $K$, at least for sufficiently large $|K|$. An
interesting question is therefore how this transition line extends
into the finite-$K$ region of the phase diagram and whether it gives
rise to a new phase.  The argument presented in Sec.~\ref{phadiagen},
which relates the gauge-field dynamics of the $N=1$ model to those in
the ordered phase of the $N>1$ models, indeed suggests that the
topological transition lines identified in the ordered phase of the
$N=2$ model should also be present in the $N=1$ model.

Let us first discuss the results of the simulations aimed at
identifying the discontinuous transition line emerging from the fully
frustrated XY model in the limit $K\to -\infty$. We performed
simulations at fixed $K=-1.5$, varying $J$, and at fixed $J=1$ and
$J=2$, varying $K$. In all three cases, we observed a discontinuous
transition. More precisely, the transitions occur at $J_{\rm
  fo}\approx 0.835$ for $K=-1.5$, at $K_{\rm fo}\approx -0.915$ for
$J=1.0$, and at $K_{\rm fo}\approx -0.684$ for $J=2.0$.

By analyzing the finite-size scaling of the maximum of the specific
heat, as in Sec.~\ref{firstorder}, we obtain the following estimates
of the latent heat: $\Delta_h\approx 0.033$ at $K=-1.5$, $\Delta_h\approx
0.21$ at $J=1.0$, and $\Delta_h\approx 0.18$ at $J=2.0$.  As already
observed for the $N=2$ model, the behavior of the latent heat is
nonmonotonic.

\begin{figure}[tbp]
  \includegraphics*[width=0.90\columnwidth]{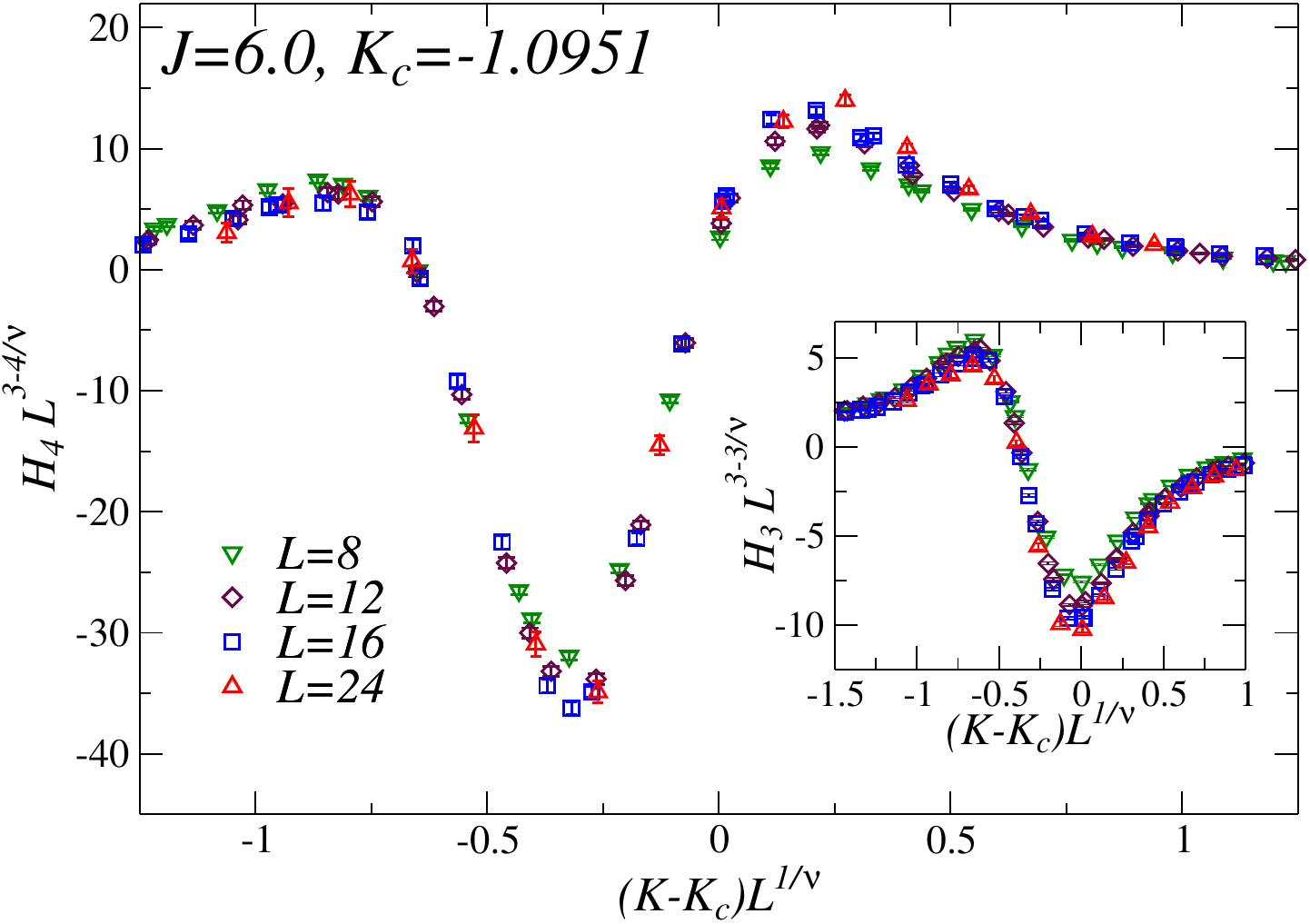}
  \includegraphics*[width=0.90\columnwidth]{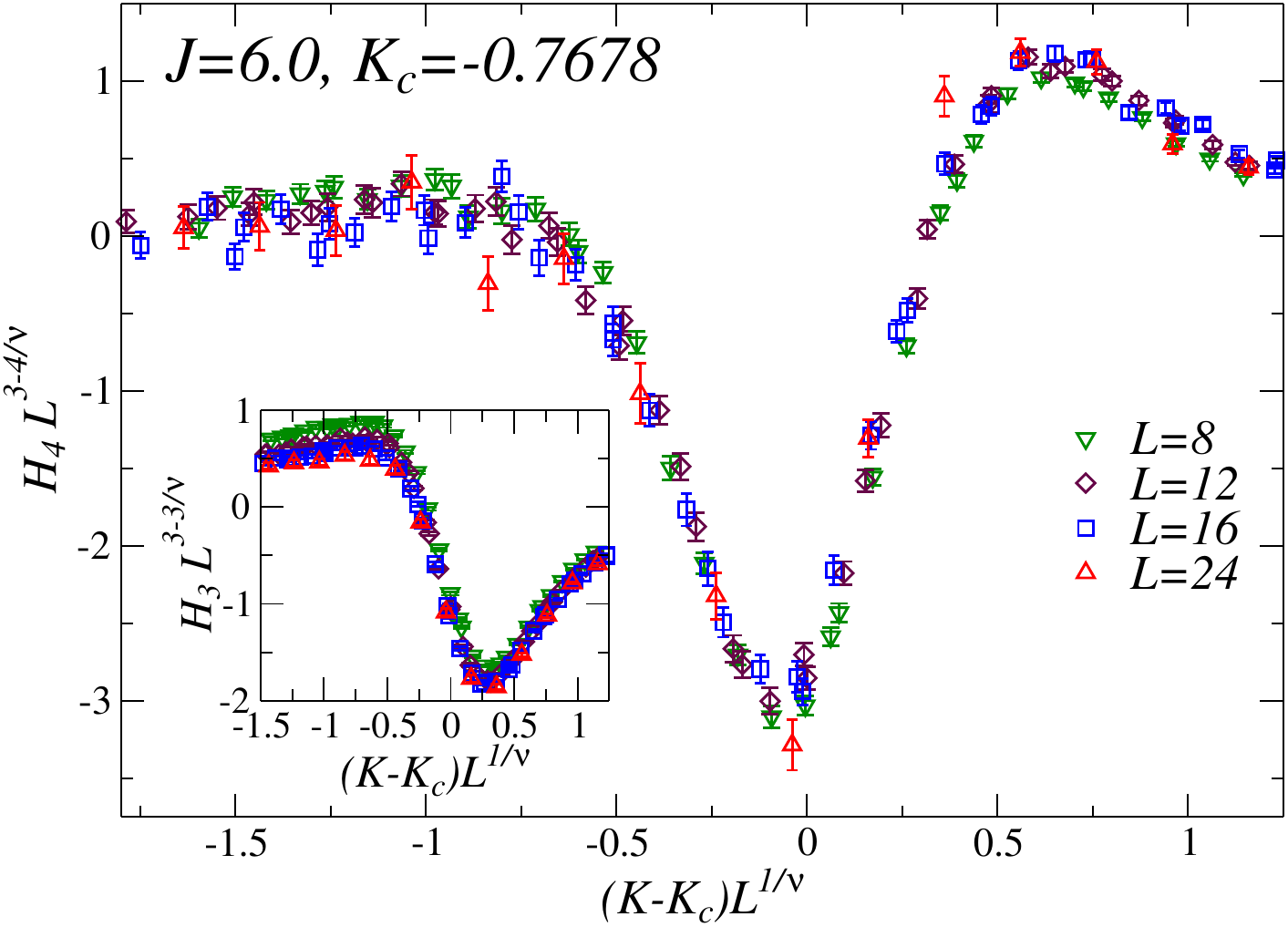}
  \caption{Scaling of the fourth (and third, in the insets) cumulants
    of the energy for the single-charge $N=1$ model at $J=6$ for
    $K\approx -1.10$ (upper panel) and for $K\approx -0.77$ (lower
    panel), obtained by fixing the critical exponent $\nu$ to the
    values of the three-dimensional Ising and XY models, respectively.
  \label{N1_ising_xy}}
\end{figure}

To investigate whether the large-$J$ transitions observed in the
ordered spin phase of the multi-component CLAH model exist also in the
$N=1$ model, we carried out simulations at $J=4$ and $J=6$, varying
$K$.  In both cases we found clear evidence of continuous transitions,
that are naturally interpreted as topological transitions, since there
is no global symmetry in the $N=1$ model that can be spontaneously
broken. FSS analyses completely analogous to those described above for
the $N=2$ case have been carried out to determine their critical
properties, using lattices of size up to $L=24$.

For $J=6$, the analysis of the third energy cumulant yields
$K_c=-1.09505(10)$ and $\nu=0.62(1)$ for the first transition (at
smaller, more negative values of $K$), and $K_c=-0.7678(3)$ and
$\nu=0.685(15)$ for the second transition. The analysis of the fourth
cumulant provides consistent results for the two transitions, namely
$K_c=-1.0951(2)$ and $\nu=0.625(15)$ for the first, and
$K_c=-0.7681(3)$ and $\nu=0.65(3)$ for the second. In this case, too,
the estimated values of the critical exponent $\nu$ are consistent
with the two transitions belonging to the Ising and XY universality
classes, respectively.  The data collapses obtained using the known
values of the critical exponent $\nu$ for the Ising and XY
universality classes are shown in Fig.~\ref{N1_ising_xy}.

A similar analysis performed for $J=4$, using the third energy
cumulant, yields the following estimates for the two transitions:
$K_c=-0.8514(1)$, $\nu=0.61(1)$, and $K_c=-0.6265(7)$,
$\nu=0.675(15)$, respectively. Using the fourth energy cumulant, we
obtain instead $K_c=-0.85145(10)$, $\nu=0.61(1)$, and
$K_c=-0.6273(5)$, $\nu=0.66(4)$.  The values of the critical exponent
$\nu$ obtained in this case for the transition at $K\approx -0.85$ are
marginally smaller than the value expected for the Ising universality
class. However, it is conceivable that $J=4$ is sufficiently close to
either a first-order transition line or a multicritical point for
significant crossover effects to occur.

\section{Conclusions}
\label{conclu}

We have presented a numerical investigation of the phase diagram and
critical behavior of 3D CLAH models with negative gauge coupling
($K<0$). These models describe charged complex spin variables ${\bm
  z}_{\bm x}$ minimally coupled to compact U(1) gauge link variables
$\lambda_{{\bm x},\mu}\in$ U(1), with the gauge coupling entering
through the gauge-invariant plaquette term in the Hamiltonian
(\ref{hamco}). Our work extends previous studies of CLAH models with
positive gauge coupling ($K>0$), whose phase diagrams and critical
properties have been extensively investigated, largely owing to their
connection with quantum U(1) gauge field theories (scalar
electrodynamics) describing the Abelian Higgs mechanism; see, e.g.,
Ref.~\cite{BPV-25} and references therein.

Negative values of $K$ favor frustration in the spin variables, as
they tend to drive the plaquette average $W$ toward negative values,
approaching $W=-1$ in the limit $K\to -\infty$, which corresponds to
maximal plaquette frustration. For $K<0$, this tendency competes with
the nearest-neighbor spin interactions with positive coupling $J>0$,
which favor ferromagnetic spin alignment. The interplay between these
competing effects leads to nontrivial physical scenarios, which turn
out to be more complex than those arising for $K>0$. In particular, we
show that this competition gives rise to a phase diagram featuring
several distinct phases, characterized not only by conventional
order-disorder transitions in the spin variables but also by different
topological phases associated with the gauge degrees of freedom.

\begin{figure}[tbp]
  \includegraphics*[width=0.90\columnwidth]{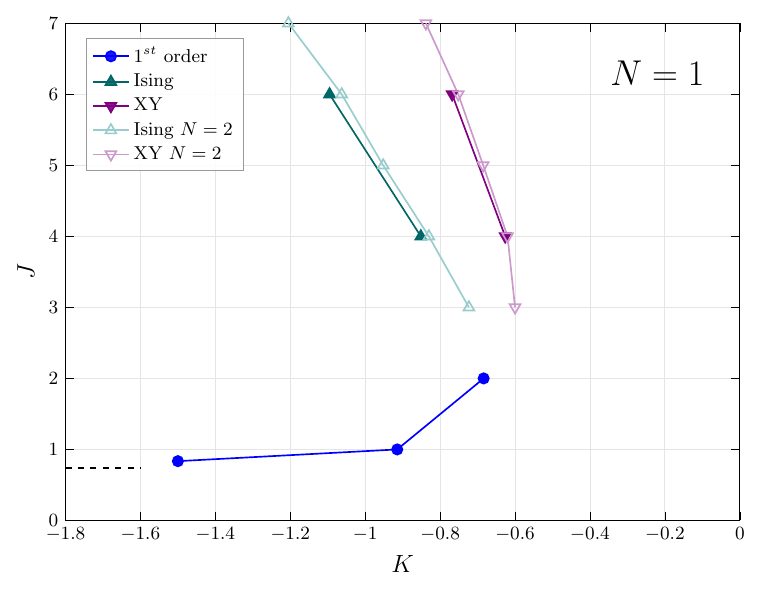}
    \includegraphics*[width=0.90\columnwidth]{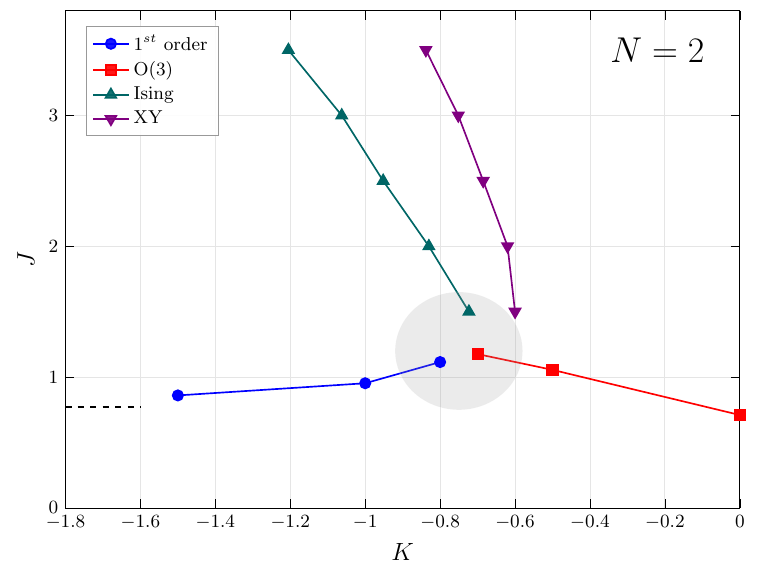}
  \caption{Sketches of the phase diagrams of the 3D CLAH models with
    negative gauge coupling $K$, for one-component (top) and
    two-component (bottom) complex scalar (unit charge) matter.  In
    the upper panel for $N=1$, the estimates $[K_c, 2J_c]$ with $K_c$
    and $J_c$ of the $N=2$ model are also reported for
    comparison. Dashed horizontal lines denote the critical value of
    the coupling $J$ in the fully-frustrated $K\to-\infty$ limit. In
    the shaded area the transition lines are expected to somehow meet,
    with the possible presence of a multicritical point.  The transition
    points for the $N=2$ model along the high-$J$ transition lines at
    values different from $J=1.5$ and $J=3$ have not been determined
    through a FSS analysis, but from the locations of the
    specific-heat maxima for a relatively small lattice size $L=10$,
    and are intended only to provide an illustration of the
    phase-diagram structure.
    \label{phadianegK}}
\end{figure}

We numerically investigate CLAH models with one-component ($N=1$) and
two-component ($N=2$) complex scalar matter fields carrying unit
charge ($q=1$). For positive gauge coupling ($K>0$), the phase
diagrams are relatively simple (see Sec.~\ref{phadiakge0}): a single
phase is present for $N=1$, and two phases for $N=2$. In the latter
case, the two phases are distinguished by the spontaneous breaking of
the SU(2) symmetry associated with the condensation of the
gauge-invariant bilinear scalar operator $Q_{\bm x}^{ab}$, cf.
Eq.~(\ref{qdef}). The phase diagrams become richer for negative gauge
coupling ($K<0$), where we find several distinct phases separated by
transition lines.

The phase diagrams obtained in this work for the single-charge CLAH
models with $N=1$ and $N=2$ are shown in Fig.~\ref{phadianegK}. The
most interesting features emerge in the large-$J$ region, where the
two cases exhibit remarkable similarities. In this regime, we identify
two additional phases that are not continuously connected to any of
the phases occurring for $K>0$. These phases appear for sufficiently
large values of both $J$ and $|K|$ and are separated by transition
lines of different nature. For $N=2$, they occur in the region where
the spin degrees of freedom condense, resulting in the spontaneous
breaking of the SU(2) symmetry. We have argued that, for sufficiently
large $J$, the behavior of CLAH models with $N>1$ in the
low-temperature ordered phase becomes essentially equivalent to that
of the $N=1$ model. We therefore expect this similarity to extend to
larger values of $N$, although dedicated numerical simulations would
be needed to verify this expectation. Finally, the upper panel of
Fig.~\ref{phadianegK} shows that the approximate relation
$J_c^{(N=1)}\approx 2J_c^{(N=2)}$, which follows from
Eq.~\eqref{ham_largeJ} upon neglecting fluctuations of $Q_{\bm x}$, is
rather well satisfied.

In particular, the phase diagram of the $N=2$ CLAH model for $K<0$
(bottom panel of Fig.~\ref{phadianegK}) exhibits two transition lines
associated with the breaking of the global SU(2) symmetry. One is a
first-order transition line that starts at $K=-\infty$, where the CLAH
model becomes equivalent to a fully frustrated O(4) vector model. As
$K$ increases, this line turns into a continuous O(3) transition line
at $K=K^*\lesssim -0.7$, which connects to the transition line found
for $K>0$ (see Fig.~\ref{phadiaN2q1}). In the large-$J$ region, two
additional ordered phases emerge, separated by transition lines that
apparently involve only the gauge degrees of freedom. For sufficiently
large $J$, within the region where the bilinear operator $Q_{\bm x}$
condenses, decreasing $K$ from the $K=0$ line first leads to a
transition line along which the transitions appear to be
continuous. The analysis of the energy cumulant is consistent with
critical behavior in the 3D XY gauge universality class~\cite{BPV-25},
characterized by the XY correlation-length critical exponent
$\nu\approx 0.6717$. After crossing an intermediate phase, we
encounter a second transition line, where the numerical results
suggest critical behavior belonging to the 3D $\mathbb{Z}_2$ gauge
universality class, characterized by the Ising correlation-length
critical exponent $\nu\approx 0.62997$.  The four transition lines
that we have identified should meet within the shaded area in the
bottom Fig.~\ref{phadianegK}.

Analogous large-$J$ transition lines are also found in the
one-component single-charge CLAH model, although in this case no
ordered phase exists, see the top Fig.~\ref{phadianegK}.  A line of
discontinuous transitions starts from $K=-\infty$, corresponding the
O(2) fully frustrated model, which runs for values of $K$ up to
$K^*\approx -0.7$.  At larger values of $J$, two additional
topological transition lines have been identified, which, together
with the discontinuous transition line, divide the phase diagram into
three distinct phases. These three transition lines likely meet at one
point, at $K^*\approx -0.7$ and $J^*\approx 2.5$.  As in the $N=2$
case, the critical length-scale exponents at these two topological
transition lines are consistent with those of the $\mathbb{Z}_2$ and
XY universality classes. This is in agreement with the theoretical
arguments predicting analogous behaviors for the $N=1$ and $N=2$ CLAH
models for sufficiently large $J$, when the $N=2$ model is within its
ordered phase where the bilinear $Q_{\bm x}$ condenses.  As discussed
in Sec.~\ref{phadiagen}, for both $N=1$ and $N=2$, the large-$J$
topological transition lines are expected to extend towards
$K\to-\infty$, likely approaching $J\to+\infty$, without intersecting
the $J=\infty$ axis at any finite value of $K$.

We emphasize that the above identifications of the universality
classes of the large-$J$ continuous transition lines are based
essentially on numerical evidence and should therefore be regarded as
conjectures. Further numerical investigations are needed to establish
the universality class of these transitions conclusively and, in
particular, to investigate the nature of the intermediate phase
between them, and that connected with the $K= -\infty$ axis 
characterized by a large degree of frustration. Moreover, it would be
interesting to understand the nature of the points where the
transition lines meet, and determine whether they exhibit peculiar
multicritical behaviors.

In conclusion, we have demonstrated that a rich phase structure
emerges from the competition between the negative gauge coupling,
which favors frustrated plaquette configurations, and the
ferromagnetic-like matter interaction, which tends to suppress
frustration. This competition gives rise to novel phases characterized
by distinct behaviors of gauge-field correlations and separated by
topological gauge transitions. These findings also motivate further
investigations of alternative formulations of lattice Abelian Higgs
models, such as those based on noncompact gauge variables (in which a
term proportional to the fourth power of the field strength must be
added to stabilize the theory for $K<0$; see, e.g.,
Ref.~\cite{BPV-21-ncAH}), as well as lattice gauge theories with
different gauge groups.

\acknowledgments

Numerical simulations have been performed using
the CSN4 cluster of the Scientific Computing Center at INFN-PISA
and the Green Data Center of the University of Pisa.

\end{document}